\documentclass[letterpaper, 10 pt, conference]{ieeeconf}  % Comment this line out
\IEEEoverridecommandlockouts                              % This command is only
\usepackage{graphicx}
\usepackage{amsmath,amssymb}
\usepackage{amsfonts,bm}
\usepackage{dsfont}
\usepackage{color}
\usepackage{cite}
\usepackage{enumerate}
\newcommand{\halfspace}{\kern 0.1em}
\newcommand{\iprime}{{i\halfspace\prime}}

\newcommand{\A}{\mathcal{A}}

\newcommand{\V}{\mathcal{V}}
\newcommand{\R}{\mathcal{R}}

\newcommand{\F}{\mathcal{F}}
\renewcommand{\O}{\mathcal{O}}
\newcommand{\Q}{\mathcal{Q}}
\newcommand{\G}{\mathcal{G}}
\newcommand{\T}{\mathcal{T}}
\renewcommand{\S}{\mathcal{S}}

\renewcommand{\Re}{\mathbb{R}}

\newcommand{\Periphery}{\mathrm{Peri}}
\newcommand{\Boundary}{\mathrm{Bndry}}
\newcommand{\Nash}{\mathrm{Nash}}

\newcommand\figref{Fig.~\ref}

\renewcommand{\emptyset}{\varnothing}

\newtheorem{definition}{Definition}
\newtheorem{condition}{Condition}
\newtheorem{remark}{Remark}

\usepackage{algorithmic}
\usepackage[ruled,vlined,linesnumbered]{algorithm2e}    % Algorithm line numbered
\newcommand{\nonl}{\renewcommand{\nl}{\let\nl\oldnl}}   % Remove line number for one line
\usepackage{listings}

\title{\LARGE \bf
Hierarchical Decompositions of Stochastic Pursuit-Evasion Games
}

\author{Yue Guan$^{1}$ 
    ~~ Mohammad Afshari$^{2}$ 
    ~~ Qifan Zhang$^{3}$ 
    ~~ Panagiotis Tsiotras$^{4}$
	\thanks{$^{1}$Yue Guan is a PhD Candidate with the School of Aerospace Engineering, Georgia Institute of Technology, Atlanta, GA, USA. Email:
		{\tt\small yguan44@gatech.edu}}%} e
	\thanks{$^{2}$Mohammad Afshari is a Postdoctoral  Fellow with the Institute for Robotics and Intelligent Machines, Georgia Institute of Technology, Atlanta, GA, USA. Email:
		{\tt\small mafshari@gatech.edu}}%}
	\thanks{$^{3}$Qifan Zhang is currently a Software Engineer at Google, Mountain View, CA, USA. This work was done while QZ was at Georgia Tech. 
	Email:
	{\tt\small qzhang410@gatech.edu}}%}
	\thanks{$^{4}$Panagiotis Tsiotras is the David \& Andrew Lewis Chair and Professor with the School of Aerospace Engineering, Georgia Institute of Technology, Atlanta, GA, USA. Email: {\tt\small tsiotras@gatech.edu}}
}

\begin{document}

\maketitle
\thispagestyle{empty}
\pagestyle{empty}

%%%%%%%%%%%%%%%%%%%%%%%%%%%%%%%%%%%%%%%%%%%%%%%%%%%%%%%%%%%%%%%%%%%%%%%%%%%%%%%%
\begin{abstract}
In this work we present a hierarchical framework for solving discrete stochastic pursuit-evasion games (PEGs) in large grid worlds.
With a partition of the grid world into superstates (e.g., ``rooms''), the proposed approach creates a two-resolution decision-making process, which consists of a set of local PEGs at the original state level and an aggregated PEG at the superstate level.
Having much smaller cardinality, both the local games and the aggregated game can be easily solved to a Nash equilibrium. 
To connect the decision-making at the two resolutions, we use the Nash values of the local PEGs as the rewards for the aggregated game.
Through numerical simulations, we show that the proposed hierarchical framework significantly reduces the computation overhead, while still maintaining a satisfactory level of performance when competing against the flat Nash policies. 
\end{abstract}

%%%%%%%%%%%%%%%%%%%%%%%%%%%%%%%%%%%%%%%%%%%%%%%%%%%%%%%%%%%%%%%%%%%%%%%%%%%%%%%%
\section{Introduction}

Pursuit-evasion games (PEGs)~\cite{ho1965differential}, also known as optimal tag games~\cite{issac-1965},  were introduced in 1960s.
An abundance of literature exists on the topic, 
and some notable examples include~\cite{Basar:1999,vidal2002probabilistic, oyler2016pursuit}.
Many of the original pursuit-evasion games are in a continuous setting, which makes their solution extremely difficult.
Often, some form of discretization is proposed to cast the problem on a finite-dimensional space.
A common numerical approach to discretize these continuous stochastic games is via a Markov chain approximation method (MCAM)~\cite{kushner2001numerical}.
MCAM approximates the original continuous stochastic PEG with a series of discrete PEGs, each formulated as a Markov game.
The discrete transition probabilities are constructed based on the stochastic differential equations (SDEs) governing the original stochastic game.
Previous work~\cite{guan2021bounded} discussed the details of the application of MCAM to PEGs.
Under certain assumptions, and as the discretization size approaches zero, the performance of the optimal policy for the discrete PEG (extended to continuous domain through holding time) converges to that of an optimal policy for the original PEG. 

To achieve a good approximation, MCAM requires a fine discretization, which leads to discrete PEGs with a prohibitively large state space. 
However, solving for the Nash equilibrium for such large games is a challenge. 
For the zero-sum case we consider in this paper, the classic value iteration for stochastic games~\cite{filar2012competitive} requires solving a matrix game (equivalent to a linear program) at \textit{each} state and at \textit{each} iteration. 
Even though efficient algorithms exist for solving linear programs (LPs)~\cite{bertsimas1997introduction}, the sheer number of LPs to be solved makes the classic ``flat" state-space approach without hierarchy extremely expensive.
To address this challenge, we propose a hierarchical framework to decompose a large PEG into multiple smaller ones.

In the single-agent domain, hierarchical decision-making has seen many successes in expediting learning~\cite{dietterich2000hierarchical}, addressing sparse rewards~\cite{levy2017learning}, etc.
One of the fundamental concepts in this area is the concept of options~\cite{sutton1999between}.
An option is a generalization of the concept of action, which, upon execution, selects primitive actions (actions from the original action space) to reach certain subgoals. 
Options can be regarded as a set of useful behaviors that an agent can use directly off-the-shelf.
By using options instead of primitive actions, an agent can shorten its planning horizon and break down complex tasks into multiple simpler subtasks.

\begin{figure}[t]
    \vspace{+5pt}
    \centering
    \includegraphics[width=\linewidth]{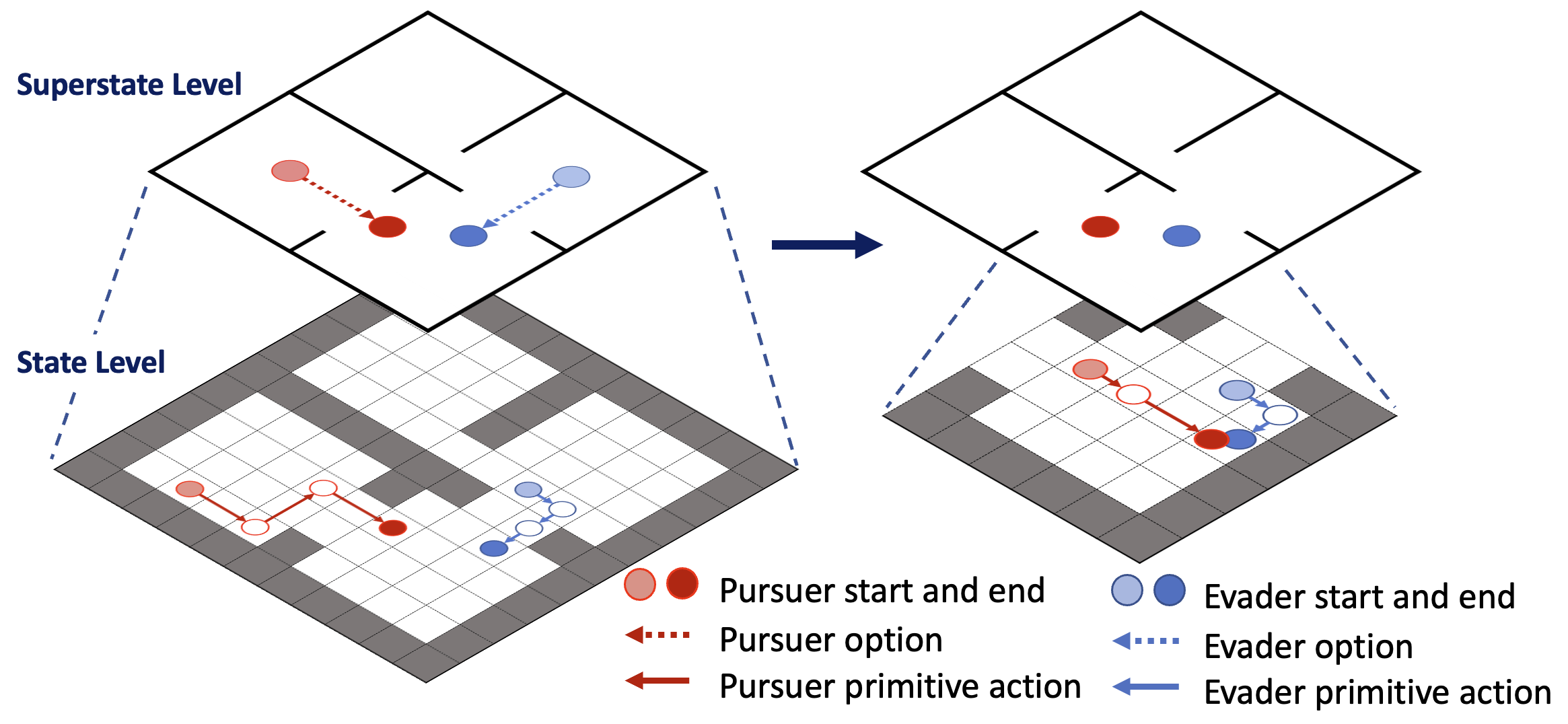}
    \vspace{-15pt}
    \caption{A schematic of the proposed hierarchical approach. \textbf{Left} presents the aggregated PEG, where the game is played at the room level and the agents select options. 
    \textbf{Right} is a local game restricted within a room.
    In this scenario, the PEG ends with a successful capture.}
    \label{fig:schematic}
    \vspace{-15pt}
\end{figure}

Previous works~\cite{ghavamzadeh2006hierarchical, yang2019hierarchical, makar2001hierarchical} have extended the option framework and the hierarchical approach to the multi-agent domain. 
However, the majority of that research focuses on cooperative games~\cite{ghavamzadeh2006hierarchical} or team games~\cite{yang2019hierarchical,makar2001hierarchical}; 
general results in the competitive setting are still lacking. 
The key challenge in the multi-agent scenario comes from the fact the subgoal exists in the joint state space of all the agents.
Subsequently, a subgoal preferable to one agent may not be favorable to the others, 
especially in competitive games. 
As a result, it is not straightforward to extend single-agent hierarchical techniques to multi-agent competitive scenarios. 

In this work, we propose a hierarchical framework to solve large two-agent zero-sum pursuit-evasion games (PEGs). 
Under this framework, the grid world is first abstracted into multiple superstates (i.e. ``rooms").
The proposed method then exploits the separable dynamics of the PEGs\halfspace\footnote{The spatial transitions of the individual agents are independent of the other agent's position and action.} to construct options that serve as generalized action to navigate agents among their \textit{individual} superstates. 
With the superstates and the options, we construct a competitive decision-making process operating at two resolutions:
\textbf{local PEGs} operating at the original resolution of the state space, each restricted within a superstate,
and an \textbf{aggregated PEG} operating at the superstate-level, where the agents select options to navigate among the superstates and the options subsequently select primitive actions to reach the target superstate.
An illustrative schematic is presented in \figref{fig:schematic}.
The local PEGs and the aggregated game have much smaller state spaces than the original game, and their Nash equilibria can be easily computed.
A key element in our approach is a way to transfer information between the two levels.
Specifically,
we connect the two resolution levels through a value propagation: we solve the Nash equilibrium of the local PEGs first and use the computed Nash values as the rewards in the aggregated PEG. 
The end result is a hierarchical policy that operates at the two resolutions.
Finally, through numerical simulations, 
we demonstrate that the proposed hierarchical approach significantly reduces the computation overhead, while still achieving satisfactory performance when competing against flat Nash policies.

%%%%%%%%%%%%%%%%%%%%%%%%%%%%%%%%%%%%%%%%%%%%%%%%%%%%%%%%%%%%%
\section{Problem Formulation}
\label{sec:Prob-form}

\begin{figure}[t]
    \centering
    \vspace{+7pt}
    \begin{minipage}[t]{0.24\textwidth}
        \centering
        \includegraphics[width=0.7\linewidth]{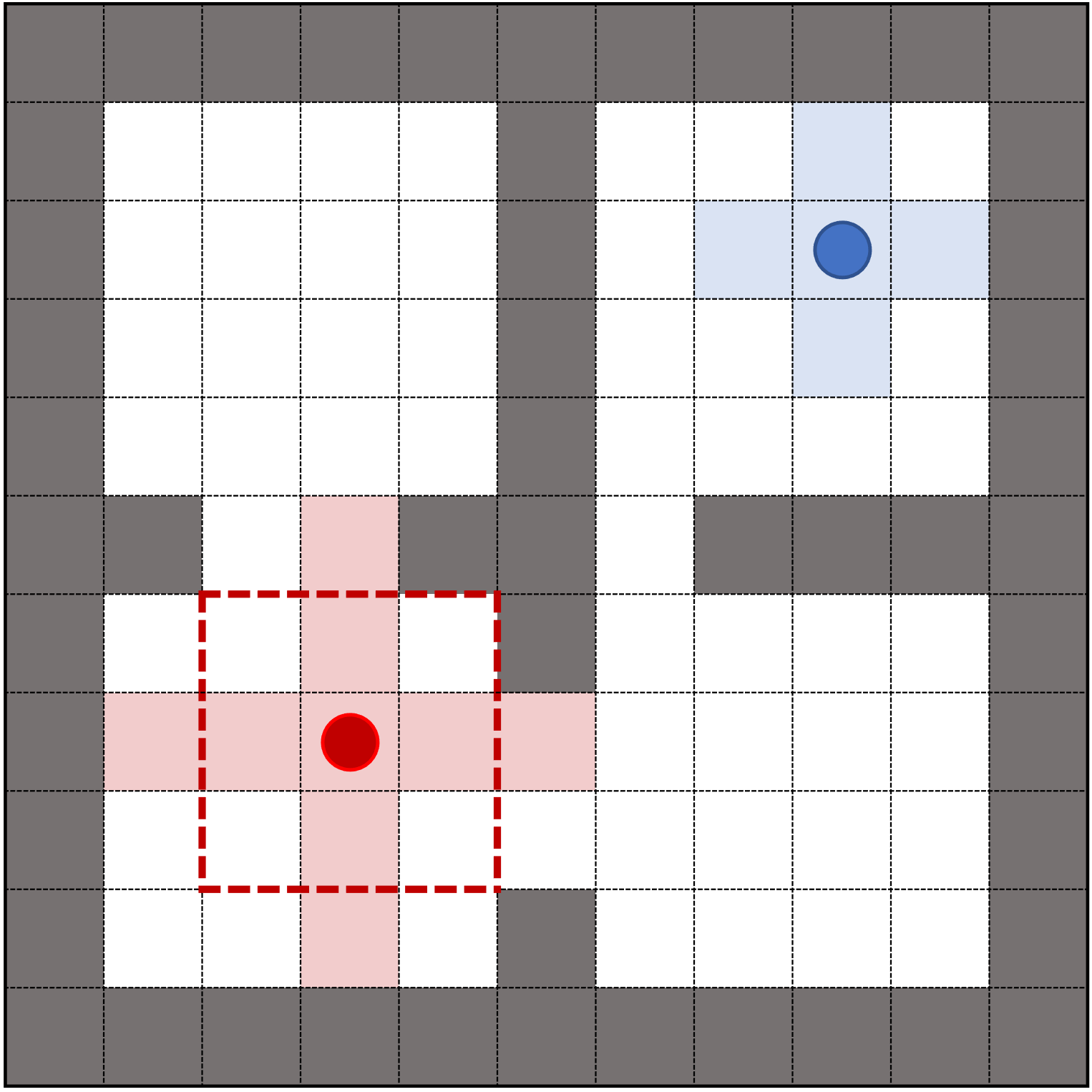}
        \vspace{-6pt}
        \caption{A pursuit-evasion game.}
        \label{fig:2x2-peg}
    \end{minipage}%
    \begin{minipage}[t]{0.24\textwidth}
        \centering
        \includegraphics[width=0.7\linewidth]{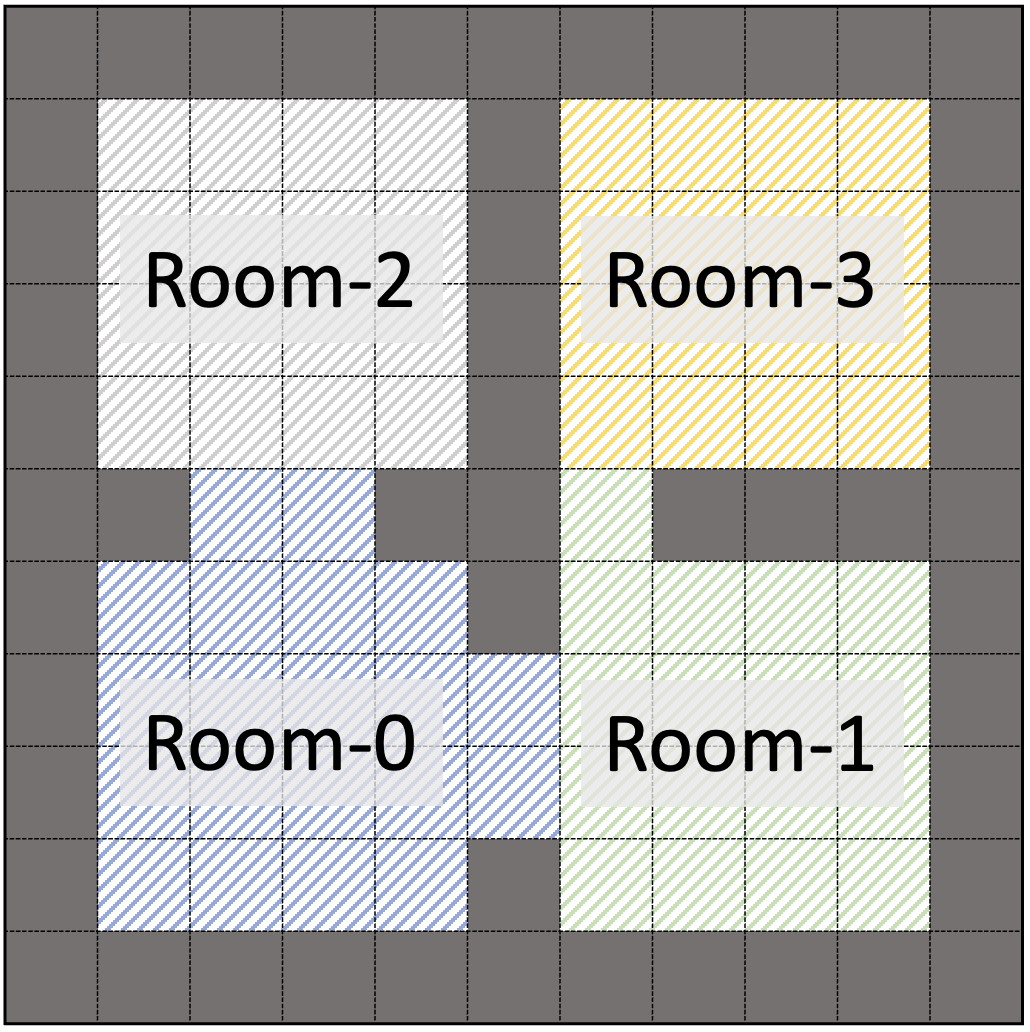}
        \vspace{-6pt}
        \caption{The four ``rooms".}
        \label{fig:rooms}
    \end{minipage}%
    \vspace{-15pt}
\end{figure}

We consider a discrete two-player pursuit-evasion game (PEG) over a grid world.
An example is presented in Fig.~\ref{fig:2x2-peg}, where a Pursuer (red) and an Evader (blue) inhabit a {11-by-11} grid world and the dark cells denote obstacles (e.g.,  walls).  
At the beginning of each episode, both the Pursuer and the Evader start at a randomly-chosen position.
The Pursuer needs to capture the Evader by having the Evader within its capture zone (red dashed square), while the Evader tries to avoid capture. 
The episode ends when the Evader is captured. 
To ensure capture, we assume that the Pursuer can move at most two cells in each of the four directions, while the Evader can move at most one cell. 
Together with the ``no-move'' action, the Pursuer has nine primitive actions, while the Evader has five.
The reachable states for the two agents are marked with the color of each agent, respectively, as shown in \figref{fig:2x2-peg}.
If an agent intends to make a move that will hit the wall, it stays at its original position and no penalty is given.

Formally, we define a discrete pursuit-evasion game (PEG) as a tuple ${\G = \langle \S^1, \S^2, \A^1, \A^2, \T^1, \T^2, \R, \F, \beta \rangle}$.
The discrete individual state space $\S^i$ corresponds to the position of agent $i$, and the discrete individual action space $\A^i$ denotes the actions agent~$i$ can take.
We refer to these actions as the \textit{primitive} actions.
We use $\S =\S^1 \times \S^2$ and $\A = \A^1\times \A^2$ to denote the joint state and action spaces.
The superscript~$i$ denotes that an entity is associated with agent~$i$. 
An entity without a superscript is assumed to be associated with the joint (state or action) space. 
%For example $\A^i$ is the action space for agent $i$, while $\A$ is the joint action space of the game.
%
The terminal set $\F$ consists of all joint states $s = (s^1, s^2)$ that correspond to successful capture.
The movement of the agents is characterized by the individual agent dynamics $\T^i: \S^i \times \A^i \times \S^i \to [0,1]$,  where $\T^i(s^{\iprime},|s^i,a^i)$ denotes the probability that agent $i$ transitions from its current 
state (position) $s^i$ to $s^{\iprime}$ under the action $a^i$.
The joint dynamics is given by $\T(s'|s,a) = \T^1(s^{1\prime}|s^1, a^1) \T^2(s^{2\prime}|s^2, a^2)$ for all states $s=(s^1,s^2) \notin \F$, and $\T(s|s,a)=1$ for all joint action $a\in \A$ if $s\in \F$.
In other words, all terminal joint states in $\F$ are absorbing.
The function $\R: \S \to [R_{\min},R_{\max}]$ provides the reward at the joint state $s$. 
We assume a zero-sum reward structure and let the Pursuer be the maximizing Agent 1 and the Evader be the minimizing Agent 2.
We further consider a sparse reward and assign $\R(s) = +1$ for $s \in \F$, and $\R(s) = 0$ otherwise.
Finally, $\beta \in (0,1)$ is the discount factor.

Assuming full observation of the joint states, we first consider Markovian mixed policies for the two agents.
Specifically, the policy~$\pi^i$ for agent~$i$ is a mapping $\pi^i: \A^i \times \S \to [0,1]$, where $\pi^i(a^i|s)$ gives the probability of choosing action~$a^i$ at the joint state $s$.
Given a policy pair $(\pi^1, \pi^2)$, we denote the induced value at each state  $s \in \S$ as
\begin{equation}
    \V^{\pi^1, \pi^2}(s) = \mathbb{E}^{\pi^1,\pi^2} \left[\sum_{t=0}^\infty \beta^t \R(s_t) \vert s_0 = s \right],
\end{equation}
where $s_t = (s^1_t, s^2_t)$ is the joint state at time step $t$.

The Nash equilibrium (NE) is then defined as a policy pair $(\pi^{1*}, \pi^{2*})$, such that, at each joint state, $s \in \S$, 
\begin{equation}
     \V^{\pi^{1}, \pi^{2*}} (s) \leq \V^{\pi^{1*}, \pi^{2*}} (s) \leq \V^{\pi^{1*}, \pi^{2}} (s) ,
\end{equation}
for all admissible policies $\pi^1$ and $\pi^2$.
That is, there is no incentive for either agent to unilaterally deviate from a NE. 
Even though multiple NEs may exist in a zero-sum game, the max-min (optimal) value of a NE is unique and can be computed via value iterations~\cite{filar2012competitive} with the following update rules:
\begin{align}\label{eqn:shapley-q-update}
    \Q_{k+1}&(s,a^1,a^2) 
    \\
    &= \R(s) + \beta \sum_{s' \in \S} \T(s'|s, a^1, a^2) \Nash\left(\bm{\Q}_{k}(s')\right),\nonumber
\end{align}
where the subscript $k$ indicates the iteration step and the Q-matrix $\bm{\Q}(s) \in \Re^{|\A^1| \times |\A^2|}$ is defined as $[\bm{\Q}(s)]_{a^1, a^2} = \Q(s,a^1,a^2)$.
At each joint state $s$, the Nash value of the Q-matrix can be computed via the following linear program from Agent 1's perspective:
\begin{equation}\label{eqn:matrix-game-nash}
    \begin{alignedat}{2}
        &\quad \quad \max_{v, \bm{\pi}^1(s)} \quad                 && v \\ 
        & \text{subject to} \quad  && v\mathds{1} - \bm{\pi}^1(s)^\top \bm{\Q}(s) \leq 0,\\
        &  \quad                    && \mathds{1}^\top \bm{\pi}^1(s) = 1, 
        ~~ \bm{\pi}^1(s) \geq 0,
    \end{alignedat}
\end{equation}
where $\bm{\pi}^1(s) \in \Re^{|\A^i|}$ is the policy of Agent 1 in vector form and $\mathds{1}$ is a column vector full of ones of compatible dimension.
The solution $v^*$ of \eqref{eqn:matrix-game-nash} gives the Nash value $v^* = \Nash \left(\bm{\Q}(s)\right)$, and $\bm{\pi}^{1*}(s)$ gives the Nash policy for Agent 1 at state $s$.
The Nash policy for Agent 2 can be solved through a similar linear program.

Although efficient algorithms exist to solve linear programs~\cite{bertsimas1997introduction}, 
there are still 4,624 states for the example shown in \figref{fig:2x2-peg} (68 position states for each agent).
That is, to compute the Nash equilibrium for this rather small problem one needs to solve 4,624 linear programs for \textit{each} value iteration. 
The number of linear programs can easily get prohibitively large.
For example, the last example in \figref{fig:test-pegs} requires almost one million LPs per iteration.
Consequently, we seek an alternative approach that will allow us to decompose a large game into smaller pieces using a hierarchical approach.

The proposed hierarchical approach is motivated by the following intuitive observations: 
(i) when the two agents are far away, knowing the approximate position of the opponent is enough to make a decent decision, 
and at this stage, the agents do not need to perform fine maneuvers to achieve satisfactory performance;
and (ii) when the two agents are close and a capture is imminent, the agents need the exact position to perform precise maneuvers to achieve (or avoid) capture.
In this case, the agents can further restrict their attention to the local environment and ignore the rest of the grid world.
Motivated by these two observations, we want to design intelligent agents that make decisions according to various resolutions of the state space, depending on the circumstances. 

\section{The Hierarchical PEG}
\label{sec:hierarchical-game}

In this section, we will first partition the grid world domain (individual state space $\S^i$) into smaller components, which we refer to as the individual superstates (or ``rooms," see Fig.~\ref{fig:rooms}).
A decision-making process is then constructed to operate at two different resolutions: 
\textbf{(i) superstate-level}, at which the two agents navigate among the ``rooms" using options~\cite{sutton1999between}, and the Pursuer tries to be in the same room as the Evader; and
\textbf{(ii) state-level}, where the two agents play a local PEG restricted within a superstate that contains a terminal state\;\footnote{With the capture region in~\figref{fig:2x2-peg}, a capture may happen when the two agents are in neighboring rooms.
For conciseness, however, we visualize local games when the two agents are in the same room.}.  
If capture occurs, the game terminates (visualized in Fig.~\ref{fig:schematic}); 
otherwise, the Evader escapes to another room, and the process reverses back to the aggregated game at the superstate level.
The end product is a hierarchical policy, whose execution will be discussed in Section~\ref{sec:hierarchical-policy}.

%%%%%%%%%%%%%%%%%%%%%%%%%%%%%%%%%%%%%%%%%%%%%%%%%%%%%%%%%%%%%
\subsection{Superstates} \label{sec:superstates}

We first formalize the definition of ``rooms" through the concept of individual superstates. 
Let $\Gamma^i = \{\gamma^i_1,\ldots,\gamma^i_\ell\}$ ($i=1,2$) be a partition of the \textit{individual} state space $\S^i$ for agent $i$, 
where each $\gamma^i_k$, $k \in \{1, 2, \dots, \ell \}$ is a subset of $\S^i$.
We say that a partition $\Gamma^i$ is an aggregated state space for agent $i$
if $\S^i  = \bigcup_{k=1}^\ell \gamma^i_k$ and 
$\gamma^i_j \bigcap \gamma^i_k = \emptyset$ for all $j \ne k$.
We refer to the subset $\gamma^i \in \Gamma^i$ as an individual \textit{superstate}, 
while the original individual state $s^i \in \S^i$ will still be referred to as an individual \textit{state} for agent $i$.
\vspace{+3pt}
\begin{remark}
The ``rooms" give a natural definition of individual superstates. 
That is, each room can be treated as a single individual superstate.
However, the concept of superstate is much more general than the concept of a ``room," as 
it does not hinge on a geometric interpretation.
% For example, in Fig.~\ref{fig:different-aggregation-size}, we have individual superstates of different sizes. 
\end{remark}
\vspace{+3pt}

Next, we identify two important classes of individual states that serve as a medium through which agents transition from one individual superstate to another.
\vspace{+3pt}

\begin{definition}[adopted from \cite{dean1995decomposition}]
The \textit{periphery} and the \textit{boundary} of an individual superstate $\gamma^i \in \Gamma^i$ are defined as
\begin{align*}
    &\Periphery(\gamma^i) \\
    &~\quad = \big\{ s^{\iprime} \notin \gamma^i \halfspace \vert \; \exists s^i \in \gamma^i, a^i \in \A^i \text{ s.t. } \T^i(s^{\iprime}|s^i, a^i)>0\big\}, \\
    &\Boundary(\gamma^i) \\
    &~\quad = \big\{ s^{\iprime} \in \gamma^i \halfspace \vert \; \exists s^i \notin \gamma^i, a^i \in \A^i \text{ s.t. } \T^i(s^{\iprime}|s^i, a^i)>0\big\}.
\end{align*}
\end{definition}

\begin{figure}[t]
    \centering
    \vspace{+5pt}
    \begin{minipage}[t]{0.23\textwidth}
        \centering
        \includegraphics[width=0.65\linewidth]{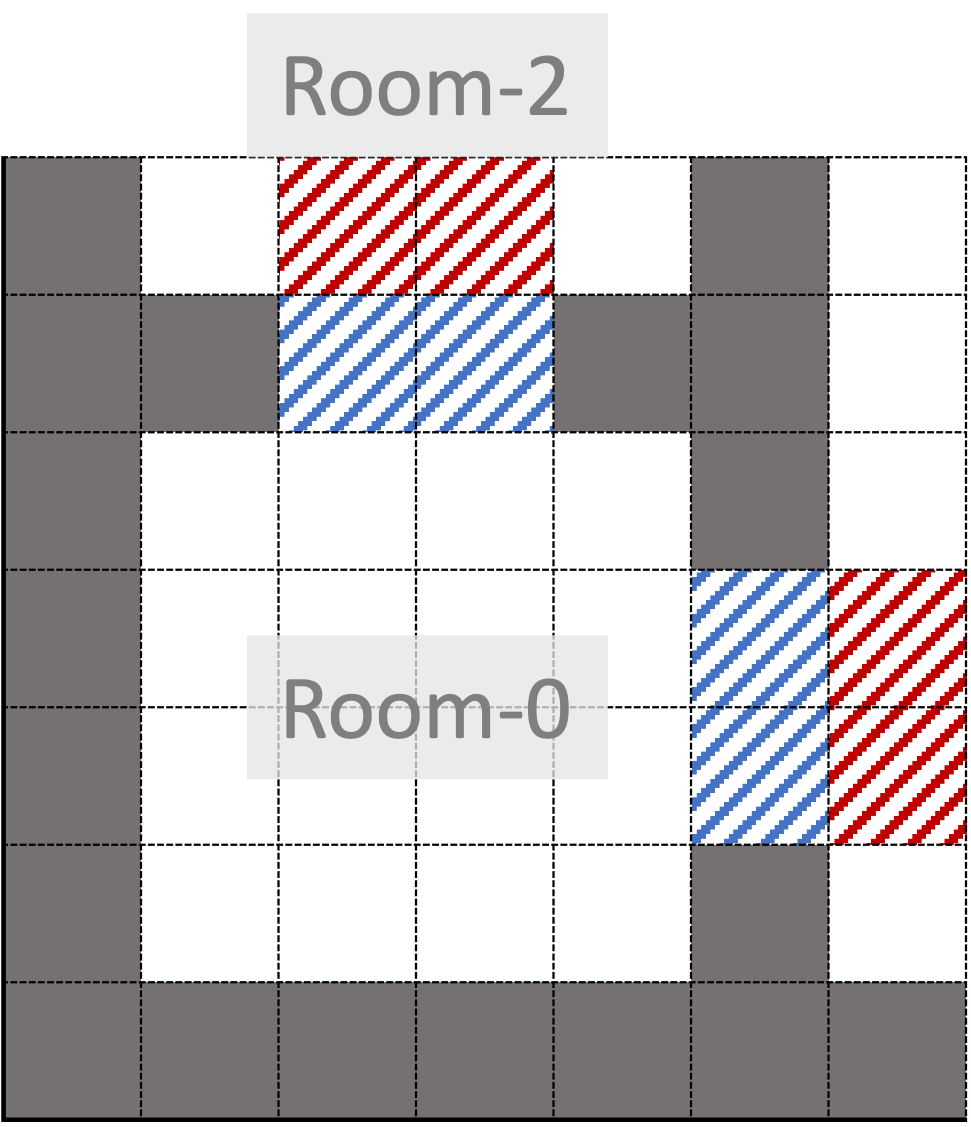}
        \vspace{-8pt}
        \caption{Boundary (blue stripes) and Periphery (red stripes) of Room-0.}
        \label{fig:boundary-periphery}
    \end{minipage}%
    \hfill
    \begin{minipage}[t]{0.23\textwidth}
        \centering
        \includegraphics[width=0.65\linewidth]{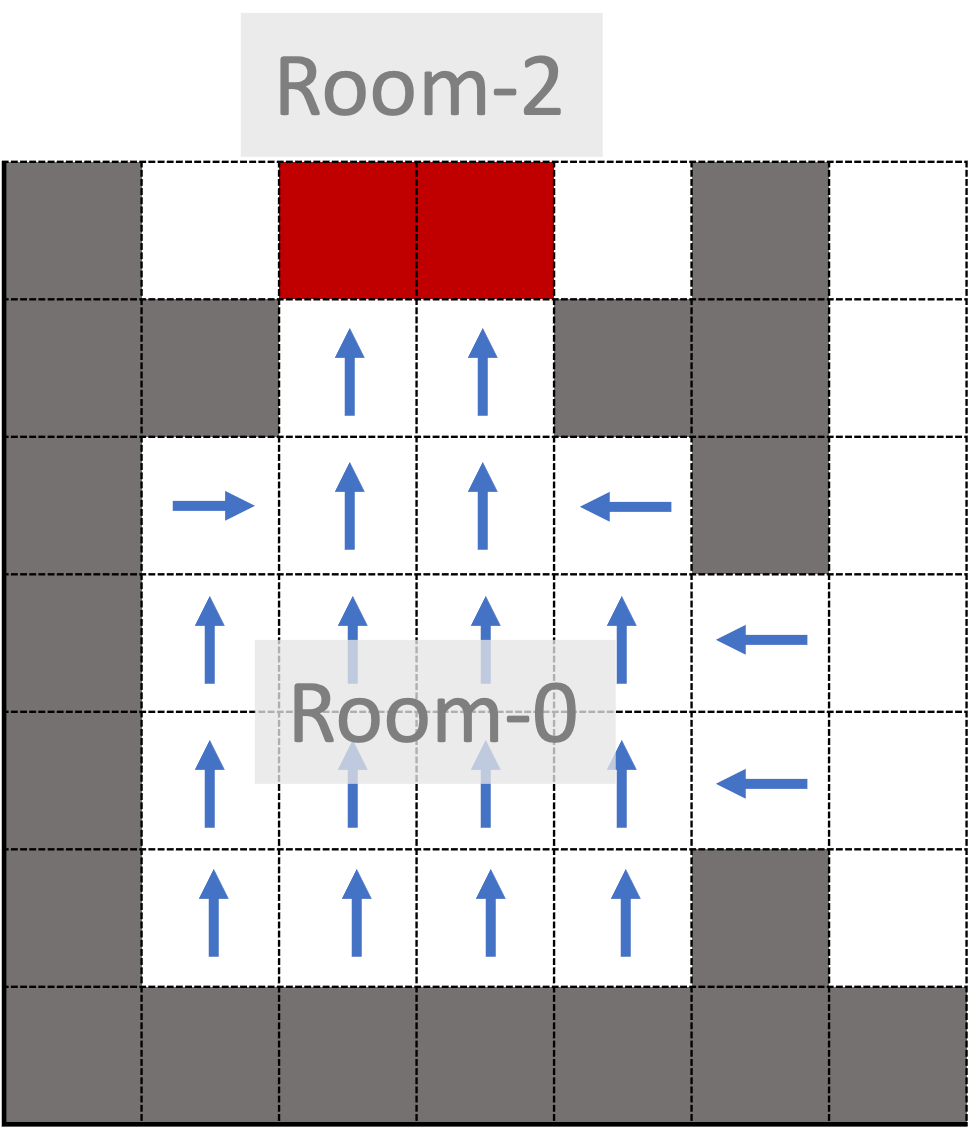}
        \vspace{-8pt}
        \caption{The option for navigating from Room-0 to Room-2.}
        \label{fig:option-example}
    \end{minipage}%
    \vspace{-10pt}
\end{figure}

\vspace{+5pt}
\noindent
In words, $\Periphery(\gamma^i)$ is the set of individual states outside $\gamma^i$ that can be reached from $\gamma^i$ within one transition.
Similarly, $\Boundary(\gamma^i)$ is the set of individual states within $\gamma^i$ that can be reached from outside $\gamma^i$ within one transition. 
The periphery and boundary for the Evader are visualized in \figref{fig:boundary-periphery}, for the environment shown in \figref{fig:rooms}.

\vspace{+5pt}

\begin{definition}
An individual superstate $\gamma^{\iprime}$ is \textit{adjacent} to $\gamma^{i}$, denoted as $\gamma^i \leadsto \gamma^{\iprime}$, if 
$\Periphery(\gamma^i) \cap \gamma^{\iprime} \ne \emptyset$.
\end{definition}
\vspace{+3pt}

The joint aggregated state space is then defined as the Cartesian product ${\Gamma = \Gamma^1 \times \Gamma^2}$, and the notion of periphery, boundary and adjacency can be easily extended to the joint superstates. \footnote{
For simplicity, we consider the case where $\Gamma^1 = \Gamma^2$.
However, the proposed framework can be easily extended to the scenarios where different abstractions of the environment are used by the two agents. 
}
For example, the periphery of a joint superstate $\gamma \in \Gamma$ is defined as
\begin{equation*}
    \Periphery(\gamma) = \big\{ s^{\prime} \notin \gamma \halfspace \vert \; \exists s \in \gamma, a \in \A \text{ s.t. } \T(s^{\prime}|s, a)>0\big\},
\end{equation*}
where $s$ and $a$ are joint state and joint action, respectively.

%%%%%%%%%%%%%%%%%%%%%%%%%%%%%%%%%%%%%%%%%%%%%%%%%%%%%%%%%%%%%
\subsection{Options}  \label{sec:options}

To construct the corresponding macro-actions of the aggregated game that result in transitions among individual superstates
we leverage the option framework~\cite{sutton1999between}.
The separable dynamics of PEGs allows us to generate options in the \textit{individual} state space, which significantly alleviates the computational burden.

\begin{definition}
An \textit{option} for agent $i$ is a tuple 
\[
o^i = \langle \S^i_{o^i}, \mathcal{F}^i_{o^i}, \pi^i_{o^i} \rangle, 
\]
where $\S^i_{o^i} \subset \S^i$ is the domain of the option, and the option $o^i$ is only available over the individual states within the domain;
$\mathcal{F}^i_{o^i}$ is the terminal set, and the option terminates once agent $i$ reaches an individual state within the set; 
finally, ${\pi^i_{o^i}: \S^i_{o^i} \to \A^i}$ is the local policy, according to which agent~$i$ selects its primitive action given its individual state within the domain.
\end{definition}

For each pair of adjacent individual superstates ${\gamma^i \leadsto \gamma^{\iprime}}$, 
we construct a Markov Decision Process (MDP) to generate the local policy for option $o^i_{\gamma^i \leadsto \gamma^\iprime}$ 
that navigates agent $i$ from individual states in $\gamma^i$ to $\gamma^{\iprime}$.
This local MDP is defined as $\mathcal{M}_{\gamma^i \leadsto \gamma^{\iprime}} = 
\langle \S^i|_{\gamma^i}, \A^i, \T^i|_{\gamma^i}, \widetilde{\R}^i_{\gamma^i \leadsto \gamma^{\iprime}}, \beta \rangle$, where
\begin{enumerate}[$i)$]
    \item $\S^i|_{\gamma^i} = \gamma^i \cup \Periphery(\gamma^i)$ is the restricted local state space;
    \item $\A^i$ is the original action space;
    \item $\T^i|_{\gamma^i}$ is the restricted transition kernel, such that, for all ${a^i\in\A^i}$,
    \begin{alignat*}{2}
        \T^i|_{\gamma^i}(s^{\iprime}|s^i, a^i) &= \T^i(s^{\iprime}|s^i, a^i) \; &&\text{if } s^i \in \gamma^i, s^{\iprime} \in \S^i|_{\gamma^i} \\
        \T^i|_{\gamma^i}(s^{i}|s^i, a^i) &= 1,  && \text{if } s^i \in \Periphery(\gamma^i).
    \end{alignat*}
    \item $\widetilde{\R}^i_{\gamma^i \leadsto \gamma^{\iprime}}: \S^i|_{\gamma^i} \to \Re$ is the local pseudo-reward, given by
    \begin{alignat*}{2}
        \widetilde{\R}^i_{\gamma^i \leadsto \gamma^{\iprime}}(s^i) &= +1, \qquad &&\text{if } s^i\in \Periphery(\gamma^i) \cap \gamma^{\iprime}, \\
        \widetilde{\R}^i_{\gamma^i \leadsto \gamma^{\iprime}}(s^i) &= 0, \qquad && \text{otherwise.}
    \end{alignat*}
\end{enumerate}
The local pseudo-reward provides an incentive for agent $i$ to move to the periphery states in the target  individual superstate $\gamma^{\iprime}$. 
Once the agent reaches the target superstate, the local MDP terminates.
The local MDP $\mathcal{M}_{\gamma^i \leadsto \gamma^{\iprime}}$ is a \textit{single-agent} problem and 
can be easily solved via value iterations or policy iterations\footnote{
% Since the option is constructed in the individual state space, these local MDPs are single-agent problems.
We let both agent maximizes with respect to this pseudo-reward, just for reaching the target superstate.
}.
The resulting optimal policy $\pi^{i*}: \S^i|_{\gamma^i} \to \A^i$ is then assigned as the local policy
for option $o^i_{\gamma^i \leadsto \gamma^\iprime}$.
Note that the local policy operates on the individual state space.
%
% \textcolor{blue}{We can further project the local policy to the joint state space, such that $\pi^{i*}(s) = \pi^{i*}(s^i)$, for all joint state $s = (s^i, s^{-i})$ such that $s^i \in \S^i\vert_{\gamma^i}$.}
% \pt{this is not clear}

We set the domain for option $o^i_{\gamma^i \leadsto \gamma^\iprime}$ as $\gamma^i \subset \S^i$, 
and the terminal set as $\Periphery(\gamma^i)$.
Consequently, agent $i$ can initiate option $o^i_{\gamma^i \leadsto \gamma^\iprime}$
within superstate $\gamma^i$ and the option automatically terminates when the agent leaves $\gamma^i$.
An example of the Evader's `Room-0 To Room-2' option is presented in Fig~\ref{fig:option-example}.
The arrows are the actions taken by the local policy at each specific cell within the room, and the red cells are the target periphery states within the individual superstate `Room-2'.

For each adjacent pair of individual superstates, we solve for all such options, and we use $\mathcal{O}^i$ to denote the set of all these options for agent $i$.
Furthermore, we let $\mathcal{O}^i(\gamma)$ denote the options available to agent $i$ at the joint superstate~$\gamma$.
% Furthermore, we define the available options for agent $i$ at a joint superstate $\gamma$ as
% $\O^i(\gamma) = \O^i(\gamma^i)$, where $\gamma = (\gamma^1,\gamma^{2})$.

%%%%%%%%%%%%%%%%%%%%%%%%%%%%%%%%%%%%%%%%%%%%%%%%%%%%%%%%%%%%%
\subsection{Local Games} \label{sec:local-games}

One key ingredient we are still missing for the aggregated game is the aggregated rewards. 
Since the aggregated game transitions to a local game when the two agents are in the same room\halfspace$^2$,
we use the Nash value of the local PEG to inform the decision-making at the superstate level.
% regarding what would happen after transitioning to playing a local game. 

Formally, we construct local games for each of the \emph{joint} superstates $\gamma$ such that $\gamma \cap \F \ne \emptyset$, where $\F$ is the set of joint states that correspond to successful capture in PEG.
% In the PEG context, these $\gamma$'s correspond to the joint superstates where the two agents are in the same room.

A local game, restricted to a joint superstate $\gamma$, is defined as a tuple $\G_\gamma=\langle \S|_\gamma, \A^1, \A^2, \T_\gamma, \R_\gamma, \beta \rangle$, where
\begin{enumerate}[$i)$]
    \item $\gamma \cap \F \ne \emptyset$.
    \item $\S|_\gamma$ is the restricted local joint state space, defined as
    \[\S|_\gamma = \gamma \cup \Periphery(\gamma).\]
    \item $\A^i$ ($i=1,2$) are the original action spaces.
    \item $\T_\gamma$ is the joint transition restricted to the local state space, given by
    \begin{alignat*}{2}
        \T_\gamma (s'|s,a^1, a^2) &= \T(s'|s, a^1,a^2), \; &&\text{if} \halfspace s \in \gamma, \halfspace s'\in \S|_\gamma \\
        \T_\gamma (s|s,a^1, a^2) &= 1, \qquad &&\text{if} \; s \in \Periphery(\gamma).
    \end{alignat*}
    \item $\R_\gamma$ is the local reward, and 
    \begin{equation*}
        \R_\gamma (s) = \R(s) \quad \forall a^i\in\A^i, \; s\in \S|_\gamma.
    \end{equation*}
\end{enumerate}
The local games terminates\halfspace\footnote{If a capture happens, the PEG terminates. Otherwise, the Evader escaped to another superstate, and we transition back to the aggregated game.}  once the agents leave the superstate $\gamma$, which corresponds to reaching the periphery states of the joint superstate $\gamma$.
Consequently, the local games have absorbing peripheries.
Meanwhile, each of these local games has a much smaller state space than the original game and can be easily solved via value iteration as in~\eqref{eqn:shapley-q-update}-\eqref{eqn:matrix-game-nash}. 
We denote the local Nash value as $\V^*_{\gamma}$ % :\S|_\gamma \to \Re$, 
and the 
local  Nash policies as $\pi^{i*}_\gamma$, $i=1,2$.
%\S|_\gamma \to \Delta_{|\A^i|}$.
The local Nash policy $\pi^{i*}_\gamma$ governs the behaviors of agent $i$ when it is in the superstate $\gamma$.
The Nash values will be used as the rewards for the aggregated game in the next subsection.

%%%%%%%%%%%%%%%%%%%%%%%%%%%%%%%%%%%%%%%%%%%%%%%%%%%%%%%%%%%%%
\subsection{Aggregated Game}    \label{sec:aggregated-game}

The aggregated game operates over the joint superstates. Instead of directly selecting a primitive action, the agent in the aggregated game selects an option based on the current joint superstate observation.
First, we define the transition probabilities between the joint superstates resulting from the options introduced in Section~\ref{sec:options}.

We leverage the discounted multi-step transition probability~\cite{sutton1999between} to properly address the different timescale on which the aggregated game operates.
With the convention $s = (s^1, s^2)$ and $s^\prime = (s^{1\prime}, s^{2\prime})$,
the discounted multi-step transition probability between two joint superstates $\gamma$ and $\gamma'$ can be computed as

\begin{align}
    \label{eqn:restricted_transitions}
    \widetilde{\T}^\beta(\gamma^{\prime}|\gamma, o^1, o^2)
    = \frac{1}{|\Boundary(\gamma)|} 
    %\Bigg[
    \sum_{s \in \Boundary(\gamma)} \hspace{-10pt}\phi(\gamma^{\prime}|s, o^1, o^2),
    %\Bigg],
\end{align}
where, 
\begin{align}
    \phi(\gamma^{\prime} |s, &o^1, o^2) = 
 %   \Bigg[ 
    \beta \sum_{s^{\prime} \in \gamma^{\prime}} 
    \T(s^{\prime}|s, \pi^1_{o^1}(s^1), \pi^2_{o^2}(s^2))
    %\Bigg] 
    \label{eqn:restricted_transition_phi}
    \\
    &+ \beta \sum_{s^{\prime} \in \gamma} \T\left(s^{\prime}|s, \pi^1_{o^1}(s^1), \pi^2_{o^2}(s^2)\right) \phi(\gamma^{\prime}|s^{\prime},o^1, o^2). \nonumber
\end{align}
The computation performed in \eqref{eqn:restricted_transitions} is equivalent to:
(a) fix agent $i$'s policy to $\pi^i_{o^i}$ ($i=1,2$) and set the $\Periphery(\gamma)$ as absorbing,
(b) start the Markov Chain induced by $(\pi^1_{o^1},\pi^2_{o^2})$ with a uniform distribution over $\Boundary(\gamma)$, 
and
(c) compute the probability of the Markov Chain ending up in joint superstate $\gamma^{\prime}$ and properly discount the probability based on the arrival time. See Appendix~\ref{appdx:discounted-transition} for more details.

% \begin{remark}
% \label{rmk:aggregated-transitions}
% A transition in the joint aggregated state space $\Gamma$ occurs when one of the individual superstate transitions. 
% \end{remark}

% \begin{remark}
% The speed advantage of the Pursuer is translated to a probabilistic advantage in the joint superstate transition. 
% That is, the joint aggregated system is more likely to make transitions where the Pursuer transitions to its target individual superstate while the Evader remains in its current individual superstate.
% \end{remark}

With the multi-step transitions, we can define the aggregated game $\G_\Gamma = \langle \Gamma^1, \Gamma^2, \O^1, \O^2, \widetilde{\T}^\beta, \widetilde{\R}, \beta \rangle$, where
\begin{enumerate}[$i)$]
    \item $\Gamma = \Gamma^1 \times \Gamma^2$ is the joint aggregated state space.
    \item $\O^i$ is the  set of options constructed for agent $i$ to navigate between its individual superstates.
    \item $\widetilde{\T}^\beta$ is the discounted transitions in~\eqref{eqn:restricted_transitions}-\eqref{eqn:restricted_transition_phi}.
    \item $\widetilde{\R}$ is the aggregated reward, given by
    \begin{alignat*}{2}
        \widetilde{\R}(\gamma) &= \frac{1}{|\Boundary(\gamma)|} \sum_{s\in \Boundary(\gamma)} \V^*_\gamma(s), \quad &&
        \text{if } \gamma \cap \F \ne \emptyset, \\
        \widetilde{\R}(\gamma) &= 0, \quad &&\text{if } \gamma \cap \F = \emptyset,
    \end{alignat*}
    where $\V^*_\gamma$ is the Nash value of the local game within~$\gamma$.
\end{enumerate}
We will denote the Nash policies for the aggregated game by $\pi^{i*}_\Gamma$  ($i=1,2$).
%\Gamma \to \Delta_{|\O^i(\gamma)|}$,
% where 
% $\O^i(\gamma)$ the options available at the joint superstate~$\gamma$.

\begin{remark}
Since the boundary states are the entry points to a superstate, when computing the aggregated transitions, we assign the initial distribution over the boundary states.
Since no prior knowledge is given regarding how agents approach this superstate, a uniform distribution is considered.
Similarly, we average the Nash value on the boundary states and assign it as the terminal reward for the aggregated game.
\end{remark}

\subsection{Proposed Algorithm}

We summarize the procedure to find the hierarchical policy for the PEG in Algorithm~\ref{alg:Hier-Decomp}.
\vspace{+5pt}
\begin{remark}
Each of the local games can be further decomposed if needed. 
The trade-off between having multiple levels of decomposition and having a single level with local games of smaller size is a topic for future investigation.
\end{remark}
\label{sec:solver-summary}
\vspace{+5pt}
\begin{algorithm}[ht]
\SetAlgoLined
\caption{Hierarchical Decomposition Algorithm}
\label{alg:Hier-Decomp}
\lstset{numbers=left, numberstyle=\tiny, stepnumber=1, numbersep=5pt}
\textbf{Inputs:} Game $\G$, Individual State Space Partition $\Gamma^i$ \;
% \vspace{+3pt}
%
\tcc{Superstates and options}
Generate the superstates ~[Section~\ref{sec:superstates}] \;
Generate option $o_{\gamma^i \leadsto \gamma^{\iprime}}$ for each agent $i$ and each adjacent individual superstate pair ~[Section~\ref{sec:options}] \;
Store the options in $\mathcal{O}^i$ \;
\tcc{Local games}
% \vspace{+2pt}
%
Generate local games for joint superstate $\gamma$ if $\gamma \cap \mathcal{F} \ne \emptyset$ ~[Section~\ref{sec:local-games}]\;
Solve the Nash equilibrium of the local games\;
Store the values of the local game within $\gamma$ as $\V^*_\gamma$ and the local Nash policies as $\pi^{i*}_{\gamma}$\;
\tcc{Aggregated game}
% \vspace{+2pt}
%
Construct the aggregated game with $\mathcal{O}^i$ as the action space and $\V^*_\gamma$ as the rewards ~[Section~\ref{sec:aggregated-game}]\;
Solve the NE of the aggregated game\;
Store the Nash policy of the aggregated game as $\pi^{i*}_\Gamma$ \;
\vspace{+3pt}
\textbf{Outputs:} Option sets $\O^i$; Local game NE $\pi^{i*}_\gamma$; Aggregated game NE $\pi^{i*}_\Gamma$.

\end{algorithm}

\subsection{Complexity Analysis}
We present a complexity comparison between the flat Nash approach and the hierarchical approach. 
Suppose we have a grid world with $M \times M$ rooms, each room consisting of $N\times N$ cells,
and we generate an individual superstate each with a single room within it.

\paragraph{Time Complexity} 
The original PEG has $(MN)^4$ joint states.
Then, for each value iteration, the flat Nash solution needs to compute $\mathcal{O}(M^4 N^4)$ matrix games.
On the other hand, we have an aggregated game with ${M^2}$ individual superstates, where each individual superstate containing a room with $N^2$ individual states.
There are at most $5M^2$ local games to be solved\halfspace$^2$, 
each with $N^4$ joint states.
There is an extra aggregated game with $M^4$ joint superstates. 
Consequently, the total number of matrix game solved per iteration\footnote{%
The hierarchical approach does not solve all these games simultaneously. 
We just sum the number of matrix games solved per iteration for each of the games to give a qualitative comparison.
}
in the hierarchical approach is 
$\mathcal{O}(M^2 N^4 + M^4)$.
% Furthermore, if one wants to minimize the number of matrix game solved, one can select $K^* = 
% \left \lfloor\sqrt[3]{{2M^2}/{N^4}} \right \rfloor$ or  $\left \lceil\sqrt[3]{{2M^2}/{N^4}} \right \rceil$.

% When $K=M$, that is, aggregating all individual states in to a single individual superstate, 
% then the game is reduced into a local game that is the original PEG, with an aggregated game with a single superstate. 
% On the other hand, with $K=1$, we will decompose the game into local games with a single state, and the aggregated game shares the same state space with the original game.
% However, in this case, the aggregated game may have a different transition probability than the original game.

\paragraph{Space Complexity} 
We present a space complexity analysis for storing the flat Nash policy and the hierarchical policies. 
The flat Nash policy for agent $i$ is a mapping from the original joint state space to a distribution over the action space $\A^i$.
Storing this flat policy is equivalent to storing a vector with $\mathcal{O}(M^4N^4|\A^i|)$ entries.

The hierarchical policy for agent $i$ has three components: the aggregated Nash policy, the local Nash policies, and the local option policies. 
The aggregated Nash policy observes the joint superstate and selects options, consequently it requires $(M^4 |\mathcal{O}^i|)$ entries to store.
We have at most $5M^2$ local games, and each has a local Nash policy that requires $(N^4|\A^i|)$ entries to store.
Finally, we have ${M}^2$ individual superstates, each with at most four adjacent superstates (``rooms"). 
That is at most $4{M}^2$ options, each requiring $(N^2|\A^i|)$ entries to store. 
Under the assumption that the number of options available at each joint superstate is of the same order as the number of actions, we have
the total amount of entries for the hierarchical policy of agent $i$ as 
$\mathcal{O}\left((M^4 + M^2 N^4 + 4M^2N^2 ) |\A^i|\right)$.

% There is an order of reduction in the space complexity, that is the memory required to store the hierarchical policy is significantly less than that of a flat Nash policy. 
% Consequently, the hierarchical policy may be desirable for intelligent agent that has limited memory onboard.

\section{The Hierarchical Policy}
\label{sec:hierarchical-policy}
The hierarchical policy constructed under the proposed framework consists of policies in two resolutions: 
(i) the aggregated Nash policy $\pi^{i*}_\Gamma$ observes the \textit{joint} superstate~$\gamma$ and selects an option ${o^i \in \O^i(\gamma)}$;
once an option $o^i$ is selected, the local policy $\pi^i_{o^i}(s^i)$ observes the \textit{individual state}~$s^i$ and selects primitive actions;
(ii) When the system transitions to a joint superstate~$\gamma$ that contains a local game, the local Nash policy $\pi^{i*}_\gamma(s)$ is deployed, and the agents directly selects its primitive action based on state observation.
If the system later leaves the superstate that contains a local game, the process reverse back to the aggregated game.
Algorithm~\ref{alg:Hier-Policy} presents the execution of the proposed hierarchical policy.
\begin{algorithm}[ht]
\SetAlgoLined
\caption{Execution of the Hierarchical Policy}
\label{alg:Hier-Policy}
\lstset{numbers=left, numberstyle=\tiny, stepnumber=1, numbersep=5pt}
\textbf{Inputs:} Option sets $\O^i$, Local game NEs $\pi^{i*}_\gamma$, Aggregated game NE $\pi^{i*}_\Gamma$, ($i=1,2$)\;
\vspace{+5pt}
\While{PEG not terminated}{
Observe current superstate $\gamma$\;
    \uIf{$\gamma$ contains a local game}{
        \While{Local PEG not terminated}{
            Observe current state $s$\;
            Select action $a^i$ according to $\pi^{i*}_{\gamma}(s)$ \;
            Environment step forward with $(a^1, a^2)$\;
        }
    }
    \uElse{
        Select option $o^i$ according to $\pi^{i*}_{\Gamma}(\gamma)$ \;
        \While{Not leaving joint $\gamma$}{
        Observe current individual state $s^i$ \;
        Select action $a^i$ according to option local policy~$\pi^{i}_{o^i}(s^i)$\;
        Environment step forward with $(a^1, a^2)$
        }
    }
}
\end{algorithm}
\noindent
The while loop in lines 5 to 9 makes the hierarchical policy non-Markovian, since once an option is selected, the option policy is executed till the option terminates.
Consequently, the primitive action taken by the agent depends not only on its current state, but also on the option it has selected when entering the superstate.
At line 12, we let the agents terminate their old options when the joint superstate makes a transition\halfspace\footnote{
Equivalent to \textit{any} one of the agents reaches the terminal set of its option.}.
Note that this ``any" condition~\cite{makar2001hierarchical} is consistent with the definition of the aggregated transitions in~\eqref{eqn:restricted_transitions}.
%
% This ``any'' termination condition is consistent with the way we define the aggregated transitions, as discussed in Remark~\ref{rmk:aggregated-transitions}.

\begin{figure}[t]
    \vspace{+5pt}
    \centering
    \includegraphics[width=0.95\linewidth]{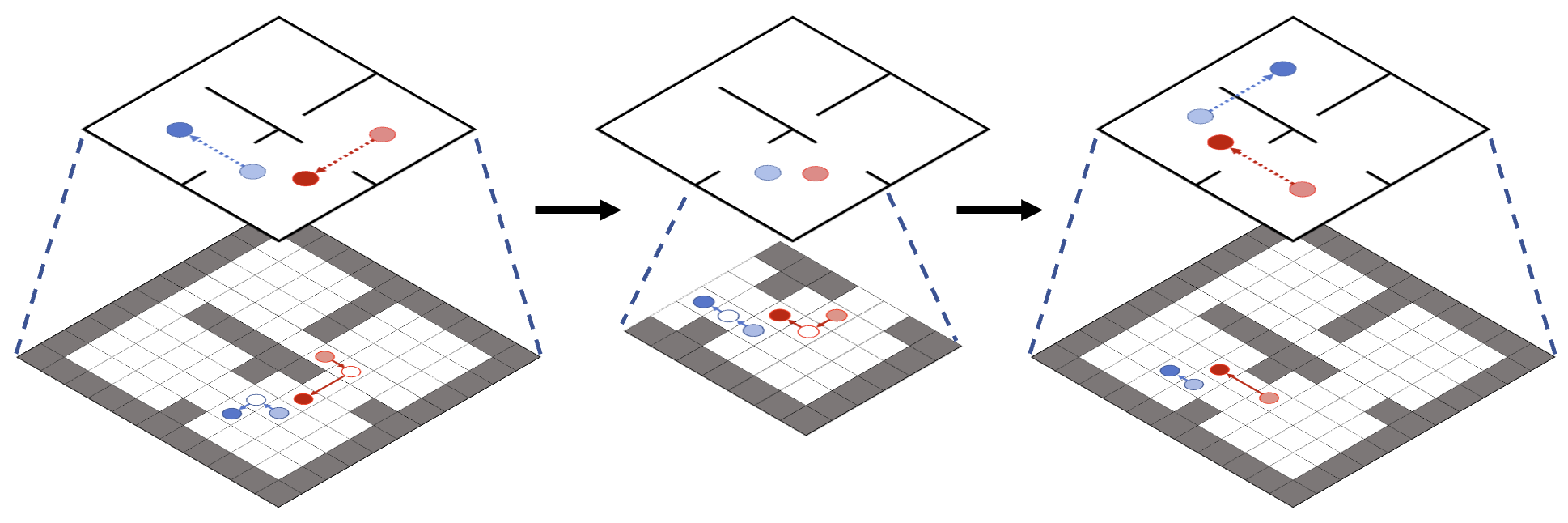}
    \vspace{-10pt}
    \caption{An example of the execution of the proposed hierarchical policy.}
    \label{fig:hierarchical-policy}
    \vspace{-10pt}
\end{figure}

An example of the execution of the proposed hierarchical policy is presented in Fig.~\ref{fig:hierarchical-policy}.
The agents first start with the aggregated game and select their options based on the current superstate. 
After the selected option executes a sequence of primitive actions, the system transitions to a local game, where the agents select primitive actions according to the local Nash. 
In the last subplot, the Evader escaped the room, and the system transitions back to the aggregated game, and the process continues.

Finally, the hierarchical policy has a different information structure from the flat policy.
The flat Nash policy requires the exact individual state information of both the agents at all time steps.
The hierarchical policy, on the other hand, requires the exact location of the opponent only when the two agents are in the same room. 
When the two agents are far away, the hierarchical policy only requires the room information of the opponent.
In that respect, the hierarchical policy can accommodate cases when the sensors of the players are inaccurate and do not provide the precise location of the other agent when the two are far apart.

\section{Numerical Results}
\label{sec:numerical}
\begin{figure}[h]
    \centering
    \hfill
    \begin{minipage}[t]{0.06\textwidth}
        \centering
        \includegraphics[width=\linewidth]{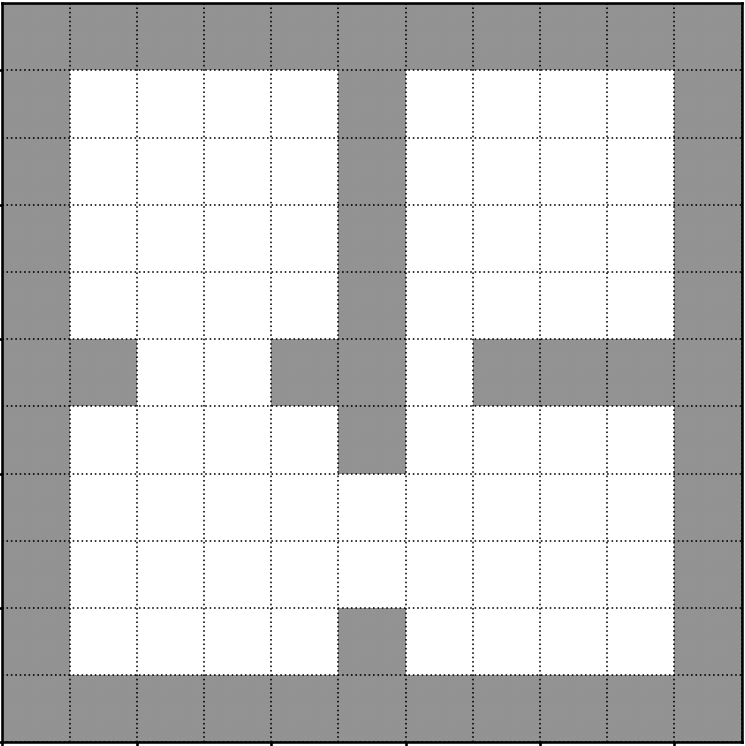}
        \vspace{-8pt}
    \end{minipage}%
    \hfill
    \begin{minipage}[t]{0.08\textwidth}
        \centering
        \includegraphics[width=\linewidth]{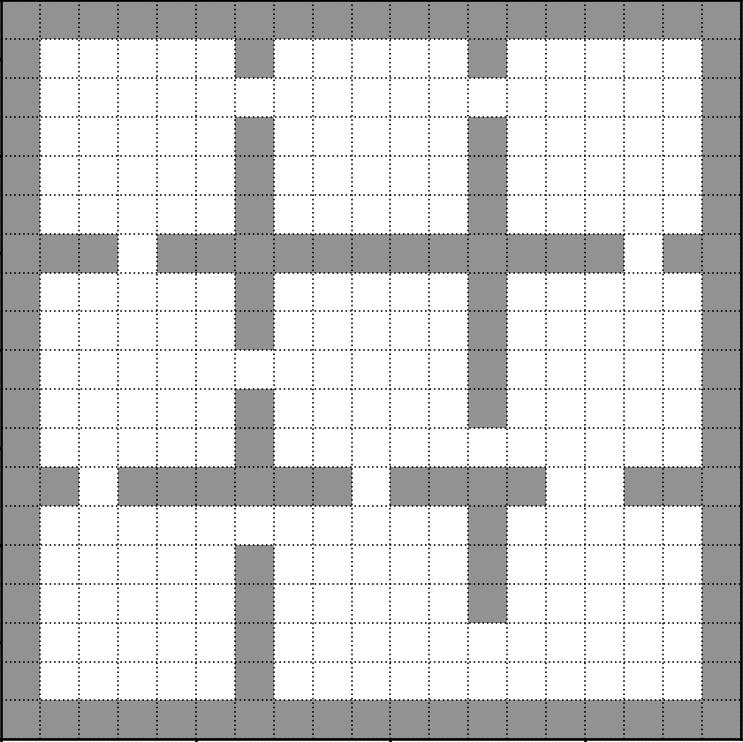}
        \vspace{-8pt}
    \end{minipage}%
    \hfill
    \begin{minipage}[t]{0.10\textwidth}
        \centering
        \includegraphics[width=\linewidth]{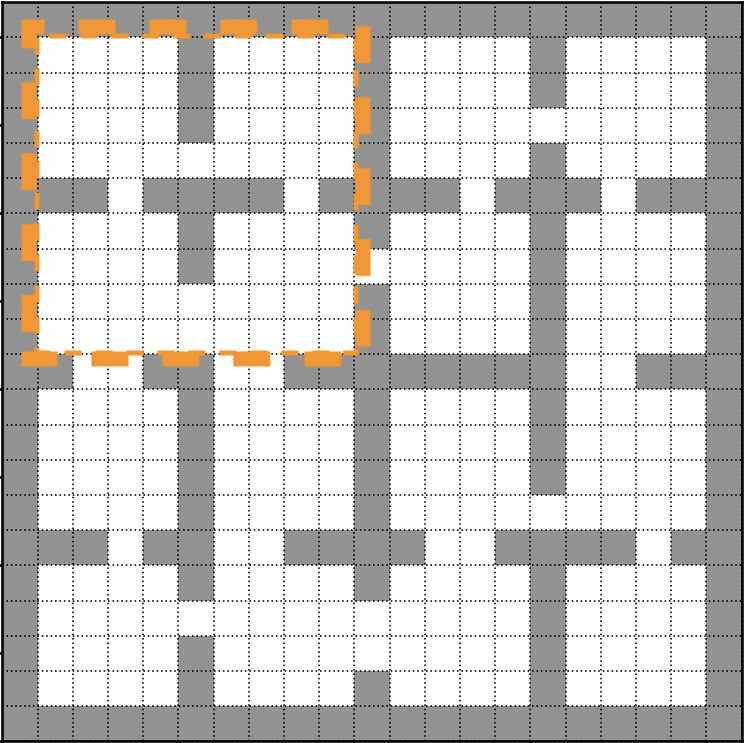}
        \vspace{-8pt}
    \end{minipage}%
    \hfill
    \begin{minipage}[t]{0.14\textwidth}
        \centering
        \includegraphics[width=\linewidth]{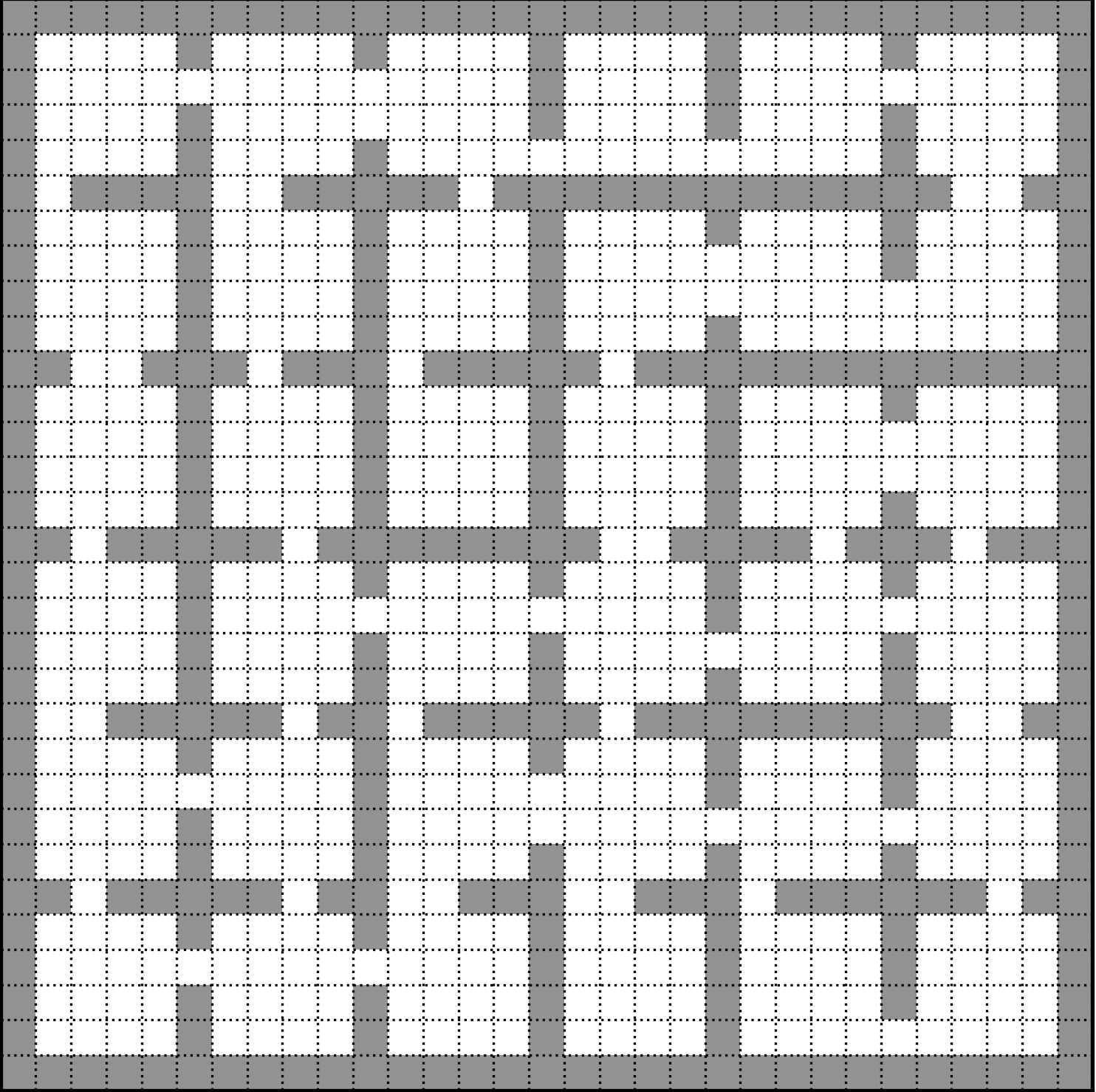}
        \vspace{-8pt}
    \end{minipage}%
    \hfill
    \vspace{-5pt}
    \caption{The test grid worlds: 2$\times$2 (4.7K joint states), 3$\times$3 (56.6K joint states), 4$\times$4 (78.9K joint states), and 6$\times$6 (0.9M joint states). }
    \label{fig:test-pegs}
\end{figure}

In this section, we verify the efficacy of the proposed algorithm through numerical simulations. 
We solve the original Nash equilibrium and the hierarchical solution for four PEGs with different sizes, shown in \figref{fig:test-pegs}.

Table~\ref{tab:hier-nash-comp-time} shows the computation time using the flat method and the proposed hierarchical approach. 
Each component of the hierarchical approach is also presented. 
One easily sees that the computation time is significantly reduced through the hierarchical decomposition of large games.
Meanwhile, the majority of the computation time is spent on the computation of NEs of the local games.
This is expected, as we have multiple local games and they are all executed at the finest resolution.
However, solving for the NE of the local games can be easily parallelized, since the local games are independent from one another.
One also observes that
when the state space is small (as in the 2$\times$2 case), the hierarchical approach is slower than the flat approach. 
This is due to the additional auxiliary tasks the hierarchical approach needs to perform (such as generating the abstract transitions and the local games) that add to the initial computational overhead.

\begin{table}[!htb]
    \caption{Computation Time Comparison [in seconds]}
    \vspace{-5pt}
    \label{tab:hier-nash-comp-time}
    \centering
	\begin{tabular}{l c c c c}
		\hline
		PEG Size \qquad      & 2 $\times$ 2 	& 3 $\times$ 3 & 4 $\times$ 4 & 6 $\times$ 6\\
		\hline
		Flat                & 380.62	& 916.85		& 7068.72   & 37412.62	    \\
		Hierarchical	    & 477.46	& 638.91		& 2095.70   & 4501.27       \\
		~- Option           & 0.92	    & 2.07			& 4.70		& 9.70	        \\
		~- Local Game	    & 462.54	& 610.98		& 2034.33	& 4257.70       \\
		~- Abstract Game	& 14.00		& 25.86			& 56.67		& 233.87
	\end{tabular}
\end{table}

\begin{figure}[ht]
    \centering
    \includegraphics[width=0.8\linewidth]{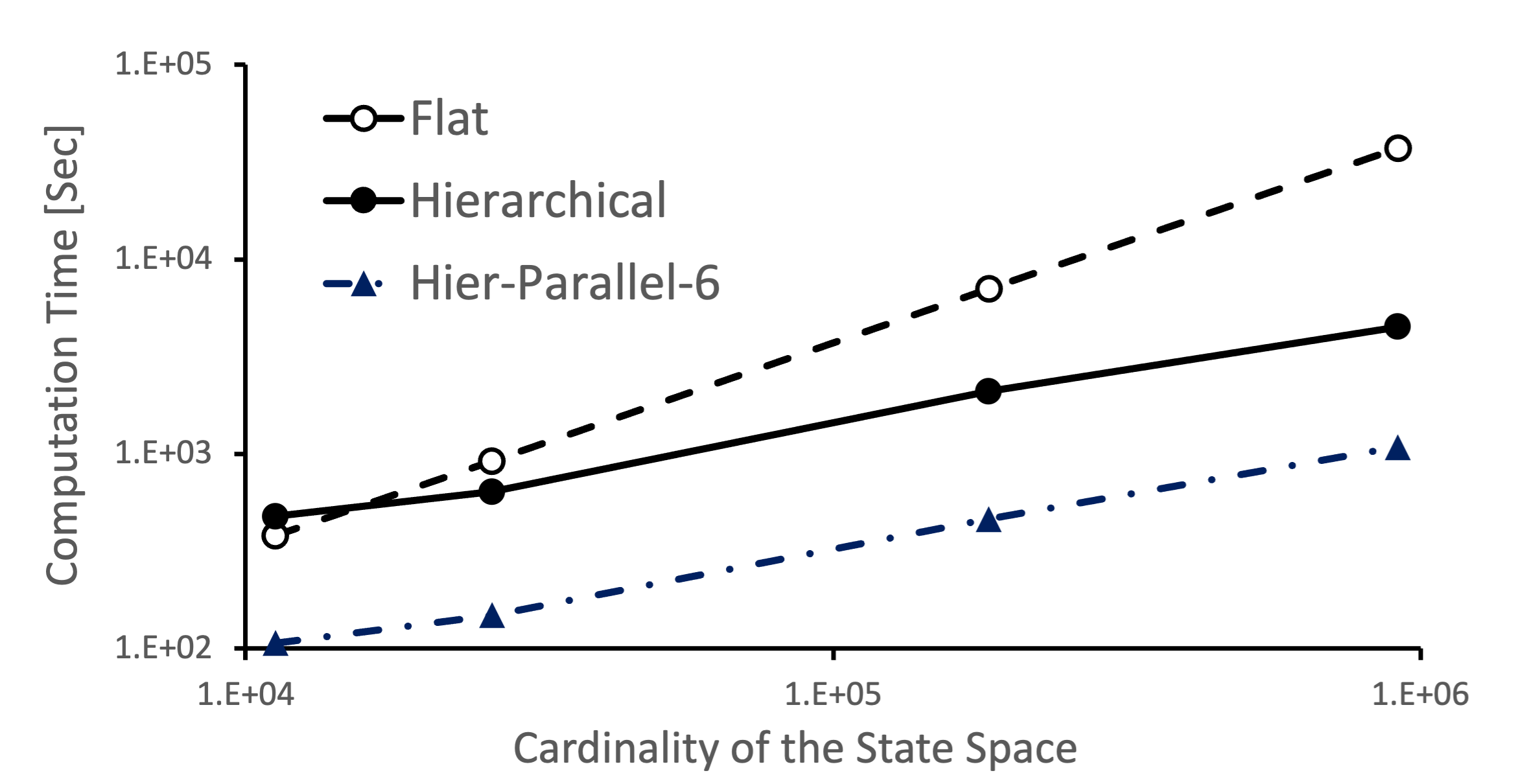}
    \vspace{-5pt}
    \caption{The log-log plot of the computation time vs. state space size trend.}
    \label{fig:comp_time_log_log}
    \vspace{-5pt}
\end{figure}

We provide a log-log plot in \figref{fig:comp_time_log_log} to better illustrate the trend of computation time vs. state space size.
The computation time of the hierarchical approach grows slower with respect to the size of the PEG than the flat approach, which confirms the motivation of this work.

% Table~\ref{tab:hier-nash-solver-called} shows the number of matrix games (LPs) solved.
% \pt{remove this table?}
% % Note that all matrix games, regardless of the original game or in the hierarchical approach (local games or the aggregated game), have roughly the same size\footnote{ 
% % Both the original and the local games have 9-by-5 Q-matrices, while for the aggregated game, the Q-matrices would be smaller as there are less number of options available at each superstate.}.
% %
% Consequently, Table~\ref{tab:hier-nash-solver-called} provides a better insight into the amount of computation saved from the decomposition, as it is independent of code implementation and the choice of LP solvers.
% However, Table~\ref{tab:hier-nash-solver-called} ignores the computation spent on constructing the local games, generating the aggregated transitions, etc.
% The additional computation time to perform these auxiliary tasks is taken into account in Table~\ref{tab:hier-nash-comp-time}.

% \begin{table}[!htb]
%     \caption{Times Matrix Game Solver Called}
%     \vspace{-5pt}
%     \label{tab:hier-nash-solver-called}
%     \centering
% 	\begin{tabular}{l c c c c}
% 		\hline
% 		PEG Size \qquad      &2 $\times$ 2 	& 3 $\times$ 3 & 4 $\times$ 4     & 6 $\times$ 6\\
% 		\hline
% 		Shapley's           & 764.0K	& 1863.3K	& 15.09M    & 82.25M    \\	
% 		Hierarchical	    & 215.2K	& 288.1K	& 0.97M	    & 2.15M	    \\
% 		~- Local Game	    & 213.7K	& 279.3K	& 0.94M		& 1.98M	    \\
% 		~- Aggregated Game	& 1.5K	    & 8.8K		& 0.03M	    & 0.17M     \\	   
% 	\end{tabular}
% \end{table}

Table~\ref{tab:hier-nash-performance} compares the performance of the hierarchical policy versus the flat Nash policy.
For each grid world, we ran 2,000 episodes, with randomly selected initial positions for the Pursuer and the Evader.
We used the average number of transitions till capture as the performance metric. 
We let the agents with different types of policies play against each other. 
When the hierarchical pursuer competed against the Nash evader, it took about 15\% more steps to capture compared to a Nash pursuer. 
This performance drop is expected, since the Nash pursuit policy is, by definition, the best response to the Nash evasion policy, but the relative performance drop decreases when the state space gets~larger.
% \footnote{%
% Only up to 6 $\times$ 6 PEG is presented, since we cannot compute the Nash equilibrium for larger PEGs.
% }.
% However, compared to the pure-pursuit and pure-evade policy constructed based on reachable sets~\cite{sun2017pursuit}, the hierarchical policy is significantly better\footnote{%
% \edit{We construct a feedback pure-pursuit policy from the reachability approach by recomputing the reachable sets every time step based on the new joint state observation. 
% Due to the present of the walls (obstacles), the reachable set is non-convex and there may have multiple time-optimal strategies for the agents. The Pursuer in this case can easily get stuck and oscillate between two time-optimal policies.}  
% }.

% We want to emphasize that when the PEG is large, or getting a solution is time sensitive, it may not be feasible to directly solve a Nash equilibrium. 
% In that case, the hierarchical approach is an alternative solution.
% By presenting the performance comparison, we want to convey the idea that the proposed hierarchical approach does produce something that is reasonable.

\begin{table}[!htb]
    \centering
    \caption{Average \# Steps till Capture}
    \vspace{-5pt}
    \label{tab:hier-nash-performance}
    \begin{tabular}{l l l l l}
    	\hline
    	P vs. E            & 2 $\times$ 2 	& 3 $\times$ 3 & 4 $\times$ 4 & 6 $\times$ 6\\
    	\hline
    	Nash vs. Nash      & 5.41		& 13.12		& 18.42	        & 25.85		\\	
    	Hier vs. Nash		            & 6.62 {\tiny{(+22\%)}}		& 15.55 {\tiny{(+18\%)}}	& 20.61 {\tiny{(+12\%)}}     & 28.18 {\tiny{(+9\%)}}  \\
    	Nash vs. Hier                   & 4.47 {\tiny{(-18\%)}}		& 10.86	{\tiny{(-17\%)}}		    & 16.26 {\tiny{(-12\%)}}			& 23.05 {\tiny{(-10\%)}}	
    \end{tabular}
\end{table}

\begin{figure}[ht]
    \vspace{+7pt}
    \centering
    \includegraphics[width=0.9\linewidth]{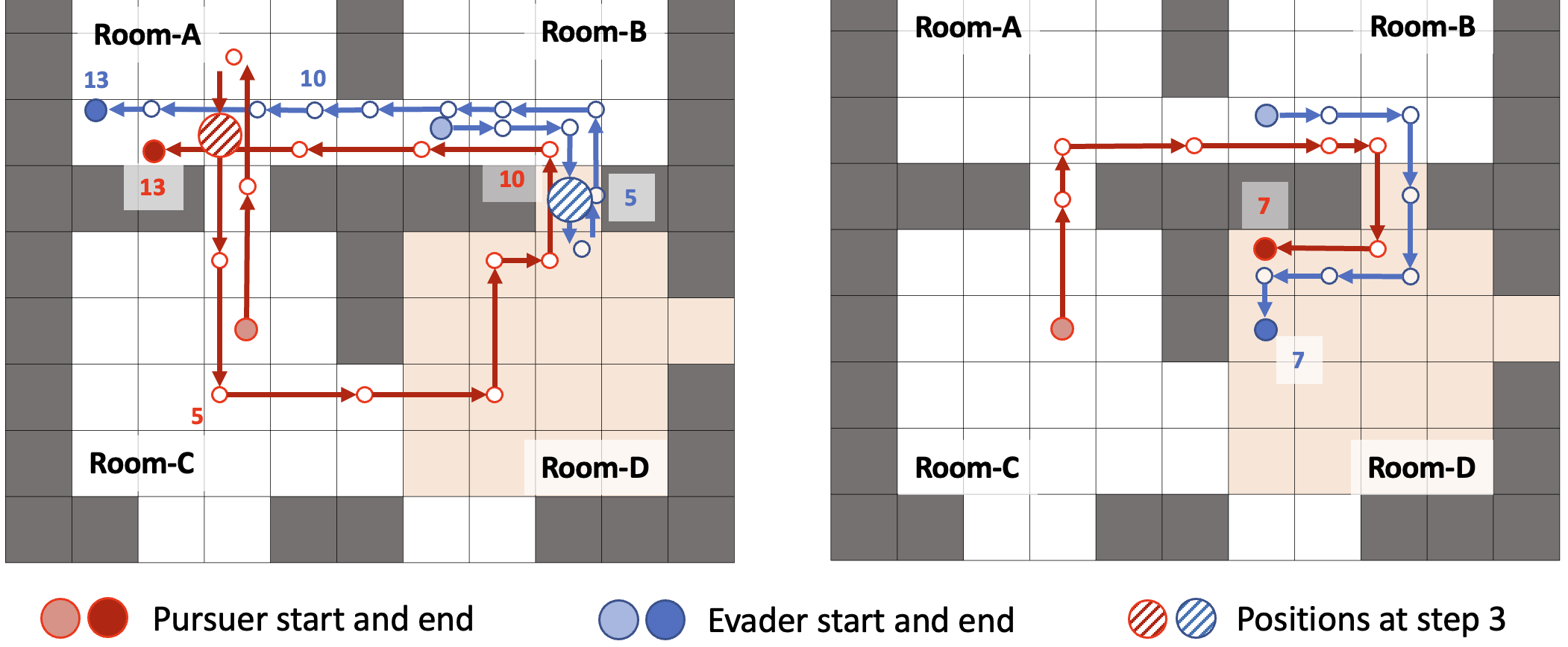}
    \vspace{-5pt}
    \caption{A comparison of sample trajectories: \textbf{left} is Hierarchical Pursuer vs. Nash Evader, and \textbf{right} is Nash Pursuer vs. Nash Evader.
    With the same starting state, the hierarchical Pursuer takes thirteen steps to capture the Nash Evader, while the Nash Pursuer only takes seven steps.}
    \label{fig:trajectory}
    \vspace{-10pt}
\end{figure}

One of the causes for the sub-optimal performance of the hierarchical approach comes from the information loss, when the agents make decisions based on the superstates.
To better illustrate this point,
we present two trajectories in \figref{fig:trajectory}.
The trajectories are truncated from the 4$\times$4 grid world, with the four rooms corresponds to the the rooms in orange in \figref{fig:test-pegs}.  
The right subplot presents a trajectory of a Nash Pursuer against a Nash Evader. 
The Nash Pursuer utilizes its speed advantage and captures the Evader with seven steps.

The left subplot in \figref{fig:trajectory} presents the trajectory of a hierarchical Pursuer against a Nash Evader.
At timestep~3 (highlighted with stripes), 
given the speed advantage, the Pursuer should directly chase down the Evader through Room-B.
However, at that moment, the hierarchical pursuer is still playing the aggregated game.
Based only on the room information, the Pursuer does not know the exact location of the Evader within room-D (light orange area).
Furthermore, if the Evader is at the bottom part of  room-D, it could easily escape to one of the other rooms. 
Consequently, the Pursuer follows the sub-optimal route down to Room-C and then come to Room-D to defend. 
Such behavior is inevitable owing to the abstraction.

Another cause for the loss of performance comes from the less frequent decision update in the hierarchical policy. 
While the Nash agent updates its `where-to-go' decision every timestep, the hierarchical agent updates its `which-room-to-go' decision only when the options terminate. Consequently, the hierarchical agents are less responsive than the flat Nash agents.

% The two causes may be addressed by doing some extra engineering. 
% For example, one may augment the information regarding the entropy point the opponent enters the room. 
% However, adding these extra mechanisms would again increase the cardinality of the state space. 
% Finding the sweet spot between the reduction of computation overhead and the satisfactory performance is obviously an interesting research direction. 

\begin{figure}[htb!]
    \centering
    \begin{minipage}[t]{0.12\textwidth}
        \centering
        \includegraphics[width=0.97\linewidth]{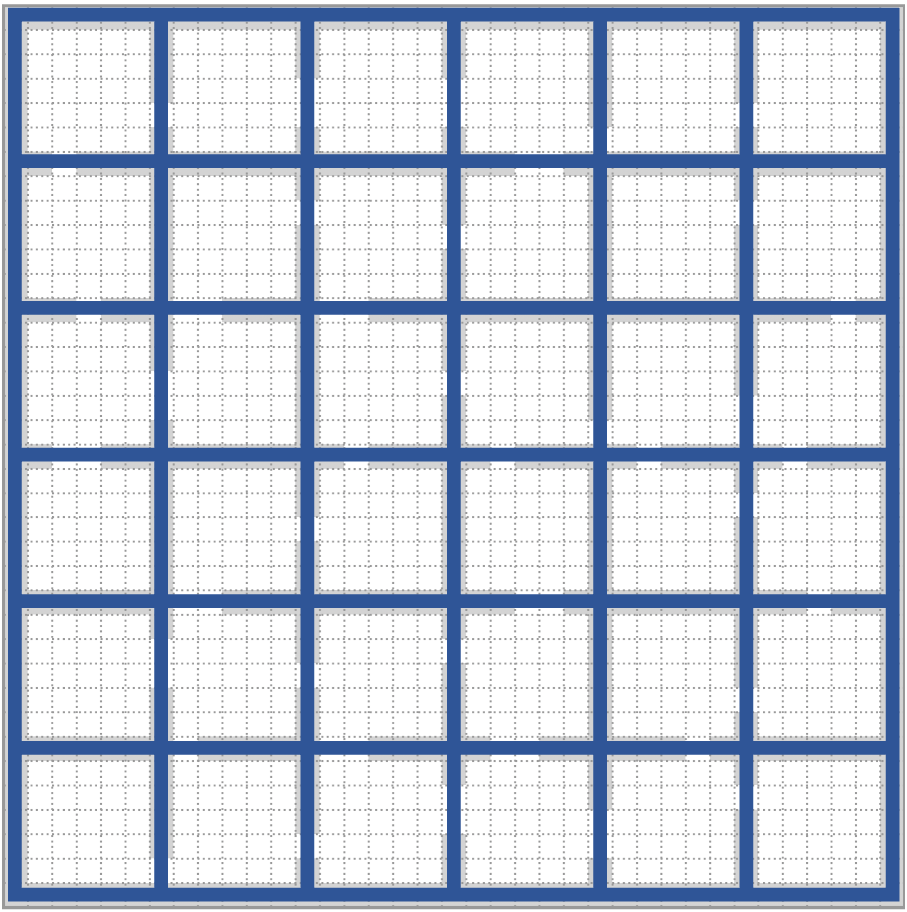}
        \vspace{-8pt}
    \end{minipage}%
    \hfill
    \begin{minipage}[t]{0.12\textwidth}
        \centering
        \includegraphics[width=0.97\linewidth]{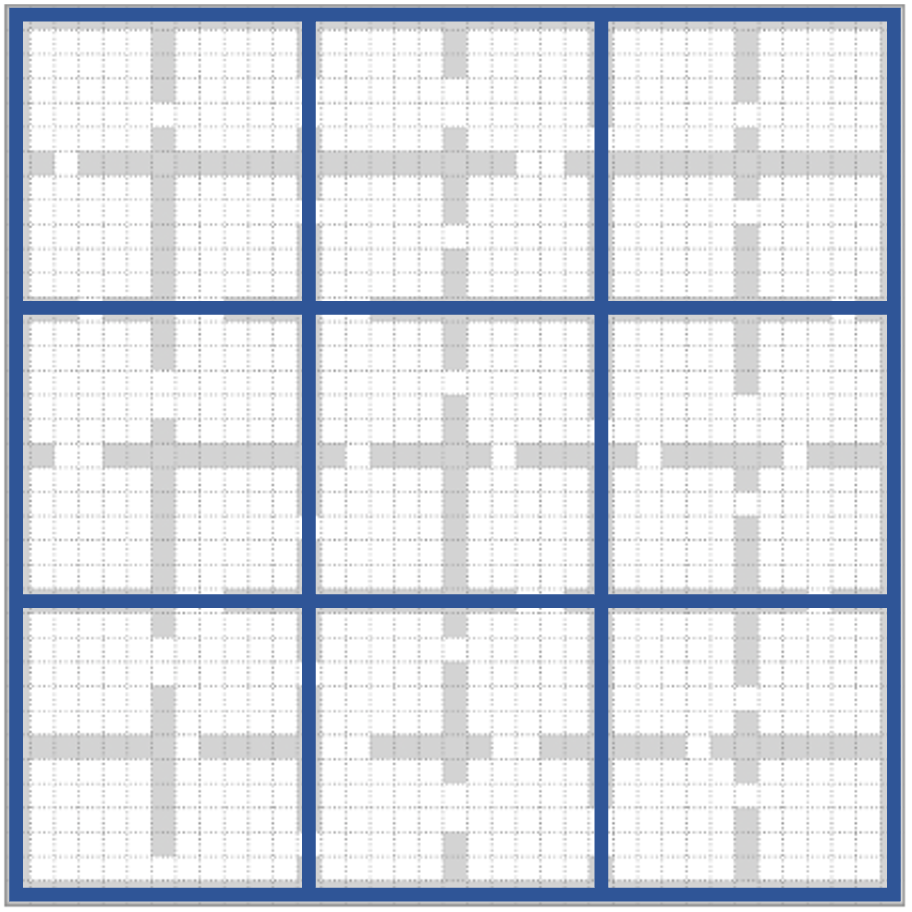}
        \vspace{-8pt}
    \end{minipage}%\hfill
    \begin{minipage}[t]{0.12\textwidth}
        \centering
        \includegraphics[width=0.97\linewidth]{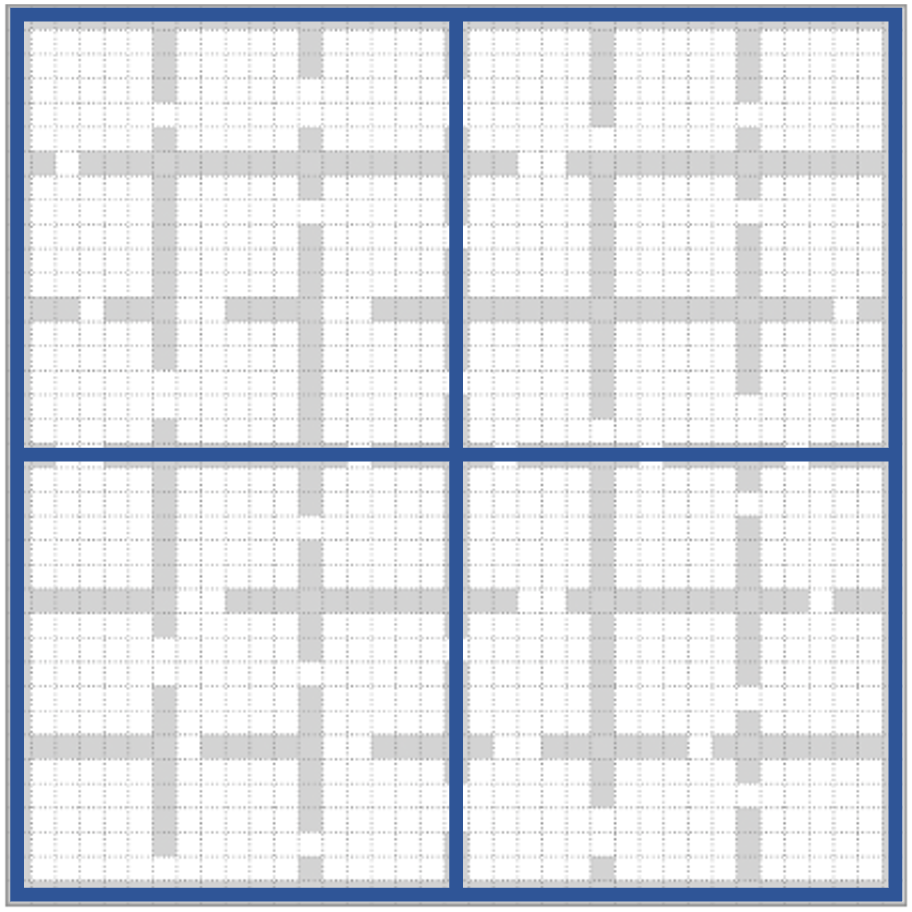}
        \vspace{-8pt}
    \end{minipage}%\hfill
    \begin{minipage}[t]{0.12\textwidth}
        \centering
        \includegraphics[width=0.97\linewidth]{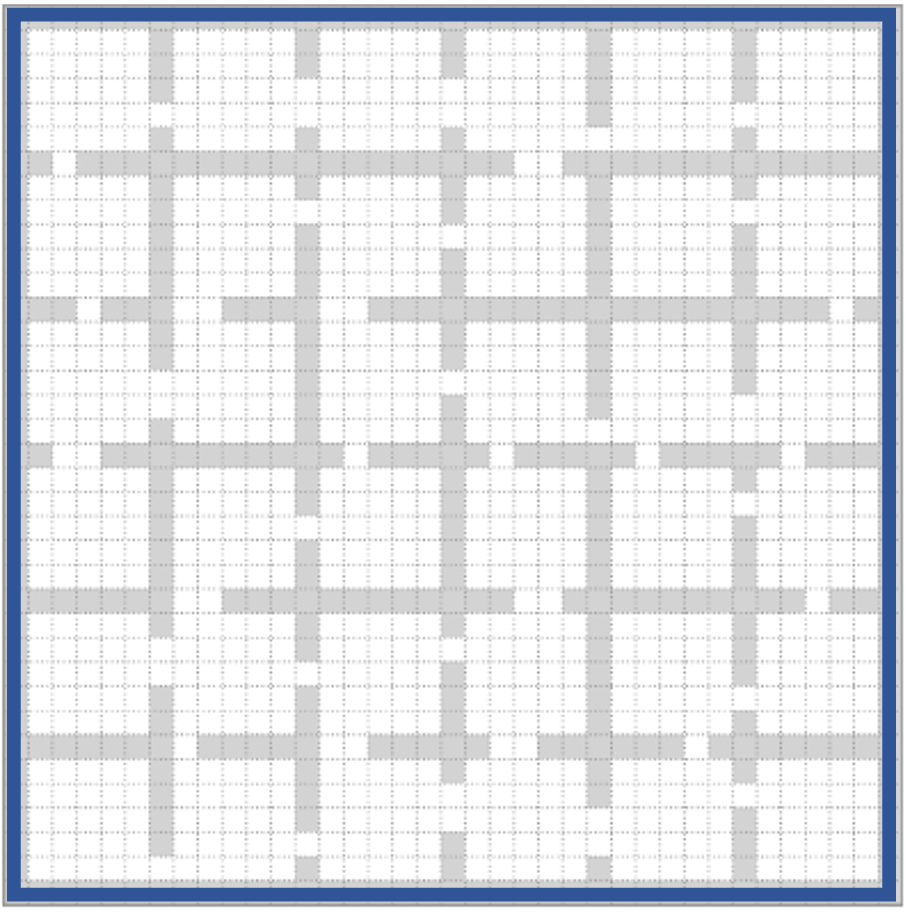}
        \vspace{-8pt}
    \end{minipage}%
    \vspace{-5pt}
    \caption{Different aggregation sizes. Each blue block is an individual superstate. From left to right: 1$\times$1, 2$\times$2, 3$\times$3 and 6$\times$6 room-blocks.}
    \label{fig:different-aggregation-size}
\end{figure}

Finally, we present the impact the aggregation size has on the algorithm. 
Instead of having one room as one individual superstate, we have a $K$$\times$$K$ block of rooms as a single individual superstate. 
An example of such multi-room aggregation for the 6$\times$6 grid world is presented in Fig.~\ref{fig:different-aggregation-size}.
The performance for different aggregation size is presented in Table~\ref{tab:different-aggregation-size}.
Note that when all rooms coalesce into a single superstate, then the original game is recovered as a single local game, and we recover the same performance as the flat Nash policy.
%\footnote{%
% The simulated average performance is slightly different between the flat Nash and the hierarchical policy with the 6$\times$6 room-block.
% This is due to the impact that the different policy execution structure has on the random number generation. 
% % The flat Nash policy directly selects the action, but the hierarchical policy needs to first consult the aggregated Nash and then pick the action according to the local Nash.   
% }. 
On the other extreme not presented here, where a single state is treated as a superstate, we can recover the original game as the aggregated game, under some additional conditions.

\begin{table}[!htb]
    \centering
    \caption{Performance with Different Aggregations}
    \vspace{-5pt}
    \label{tab:different-aggregation-size}
    \begin{tabular}{l l l l l}
    	\hline
    	       ~                        &  \multicolumn{4}{c}{each individual superstate contains} \\
    	P vs. E             & 1 room 	& 4 rooms & 9 rooms & 36 rooms \\
    	\hline
    	Nash vs. Nash                   & 25.85		& 25.85		& 25.85	    & 25.85		\\	
    	Hier vs. Nash		            & 28.18	{\tiny{(+9\%)}}	& 28.79	{\tiny{(+11\%)}}	& 30.54 {\tiny{(+18\%)}}    & 25.79   \\
    	Nash vs. Hier                   & 23.05 {\tiny{(-10\%)}}		& 21.55 {\tiny{(-16\%)}}		& 17.43	{\tiny{(-32\%)}}	& 25.82
    % 	Pure-Pursuit vs. Nash           & 8.71		& 20.04	    & 28\6.48		& 43.93
    \end{tabular}
    \vspace{-5pt}
\end{table}

When each superstate contains more states, the corresponding local games would contain a larger region of the state space and the resulting hierarchical agents are better informed in the local PEG. 
However, a larger superstate also leads to a lower resolution for the aggregated game, and the hierarchical agents ignore more information regarding the opponent's position when selecting options; 
this would result in a degraded performance due to playing an aggregated game.
The trade-off between the performance of the local games and the aggregated game is not observed in this example. 
Future work will further investigate this trade-off.

\section{Conclusion}
\label{sec:conclusion}
In this work, we proposed a hierarchical framework to decompose a large pursuit-evasion game in a grid world.
The proposed approach constructs a two-resolution decision making process, which consists of a set of local PEGs at the original state level and an aggregated PEG at the superstate level.
For the local PEGs, the agents restrict their attention within a superstate, while in the aggregated PEG, the agents utilize options to navigate among the superstates.
With this hierarchy, the decomposed PEGs have much smaller state spaces and can be easily solved to Nash equilibria. 
Through numerical simulations, we showed that the proposed approach significantly reduced the computation overhead compared to the non-hierarchical approach, while still maintaining a good level of performance. 
Future work will extend this framework to games with more than two players.
It is also of interest to investigate the theoretical bounds for the sub-optimality induced by the hierarchical decomposition.

% \hspace*{2mm}
% \textbf{Acknowledgment:} \pt{add}

\bibliographystyle{IEEEtran}
\bibliography{refs}

\clearpage
\begin{appendices}
\onecolumn
\section{Discounted Multi-step Transition Probabilities}
\label{appdx:discounted-transition}
Suppose the joint system is currently in joint superstate $\gamma$, and the agents apply a joint option $o = (o^1, o^2)$, the Q-function under the joint options is given by
\begin{equation*}
    \Q(\gamma, o^1, o^2) = \R(\gamma, o^1, o^2) + \sum_{\gamma' \in \Gamma} \sum_{\tau=1}^\infty \widetilde{\T}(\gamma', \tau| \gamma, o^1, o^2) \beta^\tau \V^*(\gamma'),
\end{equation*}
where $\widetilde{\T}(\gamma', \tau| \gamma, o^1, o^2)$ is the probability of transitioning from $\gamma$ \textit{directly} to $\gamma'$ right after $\tau$ timesteps.
In other words, $\tau$ is the timestep the joint process leaves $\gamma$ for the first time.
The discount term $\beta^\tau$ is introduced to address the different timescale on which the aggregated game operates.
With the discounted multi-step transition probability 
\begin{equation}
    \label{appdx-eqn:multi-step-agg}
    \widetilde{\T}^\beta (\gamma'| \gamma, o) = \sum_{\tau=1}^\infty
    \beta^\tau \widetilde{\T}^\beta(\gamma', \tau| \gamma, o),
\end{equation}
we can compactly write the Q-function as
\begin{equation}
    \vspace{-1pt}
    \label{appdx-eqn:Q-value-multi-step-trans}
    \Q(\gamma, o) = \R(\gamma, o) + \sum_{\gamma' \in \Gamma} \widetilde{\T}^\beta (\gamma'| \gamma, o) \V^*(\gamma').
\end{equation}
Note that the discount factor is absorbed in $\widetilde{\T}^\beta$, and therefore the discount factor is no longer explicitly presented in~\eqref{appdx-eqn:Q-value-multi-step-trans}.

Since we do not know how the agents approach the superstate $\gamma$, we assume that the agents start with a uniform distribution on the boundary of $\gamma$ (the entry points) and compute the $\tau$-step transition among superstates as 
\begin{equation*}
    \widetilde{\T}(\gamma', \tau| \gamma, o) = \frac{1}{|\Boundary(\gamma)|} \sum_{s\in \Boundary(\gamma)} \widetilde{\T}(\gamma', \tau|s,o),
\end{equation*}
where $\T(\gamma', \tau|s,o)$ is the probability of the system \textit{directly} reaching joint superstate $\gamma'$ at time $\tau$ starting from joint state~$s$. This probability can be computed as
\begin{equation*}
    \T(\gamma', \tau|s,o) = \sum_{s' \in \Periphery(\gamma) \cap \gamma'} \T(s', \tau|s,o),
\end{equation*}
where $\tau$ is the timestep the joint system leaves $\gamma$ for the first time.
Now, we can re-write the multi-step transition in~\eqref{appdx-eqn:multi-step-agg} as
\begin{align*}
    \widetilde{\T}^\beta (\gamma'| \gamma, o) &= \sum_{\tau=1}^\infty
    \beta^\tau \widetilde{\T}(\gamma', \tau| \gamma, o)\\
    % &= \frac{1}{|\Boundary(\gamma)|} \sum_{\tau=1}^\infty
    % \beta^\tau  \sum_{s\in \Boundary(\gamma)} \T(\gamma', \tau|s,o)\\
    &=\frac{1}{|\Boundary(\gamma)|} \sum_{\tau=1}^\infty
    \beta^\tau  \sum_{s\in \Boundary(\gamma)} \sum_{s' \in \Periphery(\gamma) \cap \gamma'} \T(s', \tau|s,o)\\
    &=\frac{1}{|\Boundary(\gamma)|} 
    \sum_{s\in \Boundary(\gamma)} \sum_{\tau=1}^\infty \beta^\tau   \sum_{s' \in \Periphery(\gamma) \cap \gamma'} \T(s', \tau|s,o)\\
    &= \frac{1}{|\Boundary(\gamma)|} 
    \sum_{s\in \Boundary(\gamma)} \phi(\gamma'|s,o).
\end{align*}

We further examine the term $\phi(\gamma'|s,o)$, which represents the discounted multi-step transition starting from joint state $s$ to joint superstate $\gamma'$ under joint option $o$.
\begin{subequations}
\begin{align}
    \phi(\gamma'|s,o) &= \sum_{\tau=1}^\infty \beta^\tau   \sum_{s' \in \Periphery(\gamma) \cap \gamma'} \T(s', \tau|s,o) \nonumber\\
    &=\beta \sum_{s' \in \Periphery(\gamma) \cap \gamma'} \T(s'|s,o) + \sum_{\tau=2}^{\infty}\beta^\tau \sum_{s' \in \Periphery(\gamma) \cap \gamma'} \T(s',\tau|s,o) \label{appdx-eqn:100}\\
    &=\beta \sum_{s' \in \Periphery(\gamma) \cap \gamma'} \T(s'|s,o) + \sum_{\tau=2}^{\infty}\beta^\tau \sum_{s' \in \Periphery(\gamma) \cap \gamma'} \sum_{\hat{s}\in \gamma} \T(\hat{s}|s,o) \T(s',\tau-1|\hat{s},o) \label{appdx-eqn:101}\\
    &= \beta \sum_{s' \in \Periphery(\gamma) \cap \gamma'} \T(s'|s,o) +
    \beta \sum_{\hat{s}\in \gamma} \T(\hat{s}|s,o)
    \left(\sum_{\tau=2}^{\infty}\beta^{\tau-1} \sum_{s' \in \Periphery(\gamma) \cap \gamma'}  \T(s',\tau-1|\hat{s},o)\right)\nonumber\\
    &=\beta \sum_{s' \in \Periphery(\gamma) \cap \gamma'} \T(s'|s,o) +
    \beta \sum_{\hat{s}\in \gamma} \T(\hat{s}|s,o)\phi(\gamma'|\hat{s},o). \nonumber
\end{align}
\end{subequations}
For the step from~\eqref{appdx-eqn:100} to~\eqref{appdx-eqn:101}, we used the fact that $\tau$ is the first time that the joint process leaves the superstate $\gamma$. 
Consequently, if $\tau \geq 2$, the state $\hat{s}$ at the timestep $1$ needs to be within superstate $\gamma$. 
Otherwise, the process has either already reached the target $\gamma'$ or has leaved $\gamma$ and reached superstates other than $\gamma'$. 
The rest of the derivations are simple algebraic manipulations.

In summary, we can compute the discounted multi-step transition probability starting from joint state $s$ to joint superstate~$\gamma'$ under joint option $o$ as the fixed point of
\begin{equation*}
    \label{eqn:new-iteration}
    \phi(\gamma'|s,o) = \beta \sum_{s' \in \Periphery(\gamma) \cap \gamma'} \T(s'|s,o) +
    \beta \sum_{\hat{s}\in \gamma} \T(\hat{s}|s,o)\phi(\gamma'|\hat{s},o).
\end{equation*}
The above iteration is similar to the value iteration of MDPs. 
Subsequently, with $\beta\in(0,1)$, one can easily show that the fixed point iteration above is guaranteed to converge to a unique fixed point.

Finally, the superstate-wise multi-step transition can be computed as 
\begin{equation*}
    \widetilde{\T}^\beta (\gamma'| \gamma, o) = \frac{1}{|\Boundary(\gamma)|} 
    \sum_{s\in \Boundary(\gamma)} \phi(\gamma'|s,o).
\end{equation*}

\end{appendices}

\end{document}